\documentclass[12pt]{article}
\usepackage{amssymb}
\input epsf
\makeatletter
\def\numberbysection{\@addtoreset{equation}{section}
 	\def\theequation{\thesection.\arabic{equation}}}
\makeatother

\numberbysection

\newcommand{\be}{\begin{eqnarray}}
\newcommand{\ee}{\end{eqnarray}}
\newcommand{\non}{\nonumber}
\newcommand{\id}{\mathbb{I}}
\newcommand{\tr}{\mathop{\rm tr}\nolimits}
\newcommand{\rr}{\ensuremath{\mathfrak{r}}}

\newcommand{\R}[3]{\renewcommand{\arraystretch}{0.5}
R\!\begin{array}{l}{\ \ \scriptstyle {#1}}
\\ {\scriptstyle{#2}}\\ 
{\ \ \scriptstyle{#3}}\end{array}
\renewcommand{\arraystretch}{1.0}}

\newcommand{\barR}[3]{\renewcommand{\arraystretch}{0.5}
\bar R\!\begin{array}{l}{\ \ \scriptstyle {#1}}
\\ {\scriptstyle{#2}}\\ 
{\ \ \scriptstyle{#3}}\end{array}
\renewcommand{\arraystretch}{1.0}}

\newcommand{\barRplus}[3]{\renewcommand{\arraystretch}{0.5}
\bar R^{+}\!\begin{array}{l}{\scriptstyle {#1}}
\\ {\ \ \scriptstyle{#2}}\\ 
{\scriptstyle{#3}}\end{array}
\renewcommand{\arraystretch}{1.0}}

\newcommand{\fusedS}[3]{\renewcommand{\arraystretch}{0.5}
{\tilde{\bar S}}\!\begin{array}{c}{\scriptstyle {#1}}
\\ {\scriptstyle{#2}}\\ 
{\scriptstyle{#3}}\end{array}
\renewcommand{\arraystretch}{1.0}}

\newcommand{\scriptT}[3]{\renewcommand{\arraystretch}{0.5}
\bar {\cal T}\!\begin{array}{c}{\scriptstyle {#1}}
\\ {\scriptstyle{#2}}\\ 
{\scriptstyle{#3}}\end{array}
\renewcommand{\arraystretch}{1.0}}

\newcommand{\monodromy}[2]{\renewcommand{\arraystretch}{0.5}
\bar T\!\begin{array}{c}{\scriptstyle {#1}}
\\ {\scriptstyle{#2}}
\end{array}\renewcommand{\arraystretch}{1.0}}

\newcommand{\hatmonodromy}[2]{\renewcommand{\arraystretch}{0.5}
\hat{\bar T}\!\begin{array}{c}{\scriptstyle {#1}}
\\ {\scriptstyle{#2}}
\end{array}\renewcommand{\arraystretch}{1.0}}

\newcommand{\fusedR}[3]{\renewcommand{\arraystretch}{0.5}
{\tilde{\bar R}}\!\begin{array}{l}{\scriptstyle {#1}}
\\ {\scriptstyle{#2}}\\ 
{\scriptstyle{#3}}\end{array}
\renewcommand{\arraystretch}{1.0}}

\newcommand{\fusedRplus}[3]{\renewcommand{\arraystretch}{0.5}
{\tilde{\bar R}^{+}}\!\begin{array}{r}{\scriptstyle {#1}}
\\ {\scriptstyle{#2}}\\ 
{\scriptstyle{#3}}\end{array}
\renewcommand{\arraystretch}{1.0}}

\newcommand{\scriptTfused}[3]{\renewcommand{\arraystretch}{0.5}
\tilde{\bar {\cal T}}\!\begin{array}{c}{\scriptstyle {#1}}\\ 
{\scriptstyle{#2}}\\ 
{\scriptstyle{#3}}\end{array}
\renewcommand{\arraystretch}{1.0}}

\newcommand{\fusedmonodromy}[3]{\renewcommand{\arraystretch}{0.5}
\tilde {\bar T}\!\begin{array}{c}{\scriptstyle {#1}}\\ 
{\scriptstyle{#2}}\\ 
{\scriptstyle{#3}}\end{array}
\renewcommand{\arraystretch}{1.0}}

\newcommand{\fusedhatmonodromy}[3]{\renewcommand{\arraystretch}{0.5}
\tilde {\hat{\bar T}}\!\begin{array}{c}{\scriptstyle {#1}}\\ 
{\scriptstyle{#2}}\\ 
{\scriptstyle{#3}}\end{array}
\renewcommand{\arraystretch}{1.0}}

\newcommand{\qdetR}[3]{\renewcommand{\arraystretch}{0.5}
\delta(\bar R(\theta \,, \xi_{-}))
\!\begin{array}{l}{\scriptstyle {#1}}\\ {\scriptstyle{#2}}\\ 
{\scriptstyle{#3}}\end{array}
\renewcommand{\arraystretch}{1.0}}

\newcommand{\qdetRplus}[3]{\renewcommand{\arraystretch}{0.5}
\delta(\bar R^{+}(\theta \,, \xi_{+}))
\!\begin{array}{r}{\scriptstyle {#1}}\\ 
{\scriptstyle{#2}}\\ 
{\scriptstyle{#3}}\end{array}
\renewcommand{\arraystretch}{1.0}}

\begin{document}

\begin{titlepage}
\strut\hfill UMTG--238
\vspace{.5in}
\begin{center}

\LARGE TBA boundary flows in the tricritical Ising field theory\\[1.0in]
\large Rafael I. Nepomechie\footnote{Physics Department, 
P.O. Box 248046, University of Miami, Coral Gables, FL 33124 USA}
and Changrim Ahn\footnote{Department of Physics, Ewha Womans 
University, Seoul 120-750, South Korea}
\\

\end{center}

\vspace{.5in}

\begin{abstract}
Boundary $S$ matrices for the boundary tricritical Ising field theory
(TIM), both with and without supersymmetry, have previously been
proposed.  Here we provide support for these $S$ matrices by showing
that the corresponding boundary entropies are consistent with the
expected boundary flows.  We develop the fusion procedure for
boundary RSOS models, with which we derive exact inversion identities
for the TIM. We confirm the TBA description of nonsupersymmetric
boundary flows of Lesage {\it et al.}, and we obtain corresponding
descriptions of supersymmetric boundary flows.
\end{abstract}

\end{titlepage}

\setcounter{footnote}{0}

\section{Introduction}\label{sec:intro}

A well-known (but nevertheless, remarkable) feature of integrable
quantum field theories in $1+1$ dimensions is that their exact bulk
\cite{ZZ} and boundary \cite{GZ} scattering matrices can be found. 
However, such results are generally not obtained in a systematic way
from the action; rather, one often relies on general principles
(factorizability, unitarity, crossing, bootstrap, etc.)  and educated
guesses about symmetry, mass spectrum, etc.  A case in point is the
tricritical Ising field theory -- i.e., the tricritical Ising
conformal field theory (CFT) \cite{BPZ, CFT, SCFT} perturbed by the
$\Phi_{(1\,, 3)}$ operator \cite{Za1}.  We shall refer to this field
theory as the ``tricritical Ising model '' or TIM for short.  The bulk
$S$ matrix was proposed in \cite{Za2}, and boundary $S$ matrices were
proposed in \cite{Ch, Ne1}.  This field theory has several notable
properties, which render it a very attractive toy model: it is
unitary; it is supersymmetric; and it is one of the simplest examples
of a model of massive kinks, whose scattering matrices are of RSOS
\cite{Ba, ABF} type.  Moreover, the bulk and boundary soliton $S$ matrices
\cite{ABL, AK} of the $N=1$ supersymmetric sine-Gordon model \cite{sg,
Ne2} contain the corresponding TIM $S$ matrices as one of the factors.
 
A thermodynamic Bethe Ansatz (TBA) analysis \cite{YY} can provide a
nontrivial check on a given bulk \cite{AlZa1, AlZa2} or boundary
\cite{LMSS, DRTW, AN} scattering matrix.  Indeed, the $S$ matrices
serve as the input of the ``TBA machinery,'' whose output consists of
certain data (central charge \cite{BPZ, CFT}, boundary entropy
\cite{Ca, AL}) which characterizes the corresponding CFT. For the TIM,
a TBA check of the proposed bulk $S$ matrix \cite{Za2} was performed
in \cite{AlZa2}.

One of the principal aims of this paper is to perform an analogous TBA
check of the boundary $S$ matrices which have been proposed in
\cite{Ch, Ne1}.  Such an analysis is technically nontrivial, since
neither the bulk nor boundary $S$ matrices are diagonal.  As in the
bulk case \cite{AlZa2}, the key step is the derivation of an exact
inversion identity which is obeyed by an appropriate transfer matrix. 
For the boundary case considered here, the transfer matrix is of the
``double-row'' type \cite{Sk}.

A second aim of this paper is to develop the techniques for deriving
the necessary inversion identity.  We do this in an extended appendix,
building on earlier work on fusion for vertex \cite{vertex, KS, MN}
and RSOS \cite{RSOS, BPO, Zh} models.  The main idea is to formulate
an RSOS open-chain fusion formula, and to show that the TIM fused
transfer matrix is proportional to the identity matrix.

A third aim of this paper is to derive TBA descriptions of TIM
massless boundary flows.  Let us recall \cite{Ca, Ch, Ne1} that the
tricritical Ising CFT has a discrete set of (super) conformal boundary
conditions.  Boundary perturbations can lead to flows among these
boundary conditions \cite{Ch, LSS, Af, GRW, FPR, Ne1}.  A TBA
description of the nonsupersymmetric flows was proposed in \cite{LSS}
on the basis of an analogy with the Kondo problem.  Here we give a
derivation of that TBA result, as well as the results for
supersymmetric flows not considered in \cite{LSS}, directly from the
TIM scattering theory.  (An alternative approach based on a lattice
formulation of the TIM is considered in \cite{FPR}. However, it seems
that this approach cannot generate the boundary entropies.)

The outline of this article is as follows.  In Sec.  \ref{sec:TIM}, we
review the bulk and boundary $S$ matrices \cite{Za2, Ch, Ne1} which
will serve as inputs for our TBA calculation.  There are two boundary
$S$ matrices that are not supersymmetric; and there are two boundary
$S$ matrices which do have supersymmetry, which we call NS and R. We
also briefly review the classification of (super) conformal boundary
conditions, certain pairs of which are connected by boundary flows.
In Sec.  \ref{sec:yangtransfer}, we carry out the first step of the
TBA program, which consists of constructing the so-called Yang matrix
\cite{Ya} and relating it to a commuting transfer matrix.  For the
problem at hand, we require a boundary RSOS version of the Yang
matrix, which is an interesting generalization of the known case of
periodic boundary conditions.  In Sec.  \ref{sec:inversion}, we use an
exact inversion identity to determine the eigenvalues of the transfer
matrix in terms of roots of certain Bethe Ansatz equations. We 
restrict our attention here to the NS case. In Sec. \ref{sec:TBA} we 
use these results to derive the TBA equations and boundary entropy.
Moreover, we find massless scaling limits which correspond to
boundary flows, both for the NS case and the nonsupersymmetric cases.
In Sec. \ref{sec:Rcase} we briefly discuss the R case, which is closely 
related (in fact, dual) to the NS case. Our conclusions are presented
in Sec. \ref{sec:conclusion}. In an Appendix, we give a brief
account of the fusion procedure for RSOS models with boundary, and
provide the derivation of the TIM inversion identity.

\section{TIM scattering theory}\label{sec:TIM}

We briefly review in this Section some pertinent results on the TIM
scattering theory.  We first define the bulk model as a perturbed bulk
CFT, and give the bulk $S$ matrix \cite{Za2}.  We then enumerate the
possible (super) conformal boundary conditions, and give the boundary
$S$ matrices which have been proposed \cite{Ch, Ne1} to describe
certain perturbations of some of these boundary conditions.  Two of
the boundary $S$ matrices do not have supersymmetry, and two of them
do.  Many of the notations used in this paper are introduced in this
Section.

\subsection{Bulk}

The bulk TIM is defined by the ``action'' \cite{Za2}
\be
A = A_{{\cal M}(4/5)} + \lambda \int_{-\infty}^{\infty} dy 
\int_{-\infty}^{\infty} dx\  
\Phi_{({3\over 5}\,, {3\over 5})}(x \,, y) \,, \qquad \lambda < 0 \,, 
\label{bulkaction}
\ee
where $A_{{\cal M}(4/5)}$ is the action for the tricritical Ising CFT
(i.e., the minimal unitary model ${\cal M}(4/5)$ with central charge
$c={7\over 10}$), and $\Phi_{({3\over 5}\,,{3\over 5})}$ is the
spinless $(1\,, 3)$ primary field of this CFT with dimensions
$({3\over 5} \,, {3\over 5})$.  Moreover, $\lambda$ is a bulk
parameter with dimension length${}^{-{4\over 5}}$.  We restrict our
attention to the case $\lambda < 0$, for which there is a three-fold
vacuum degeneracy, and the spectrum consists of massive (mass $m >0$)
kinks $K_{a\,, b}(\theta)$ that separate neighboring vacua, 
$a \,, b \in \{ -1\,, 0 \,, 1\}$ with $|a - b|=1$.  Multi-kink states
\be
K_{a_{1} \,, b_{1}}(\theta_{1})\ 
K_{a_{2} \,, b_{2}}(\theta_{2}) \ldots \non 
\ee 
must obey the adjacency conditions $b_{1} = a_{2}\,, $ etc. 

\begin{figure}[tb]
	\centering
	\epsfxsize=0.25\textwidth\epsfbox{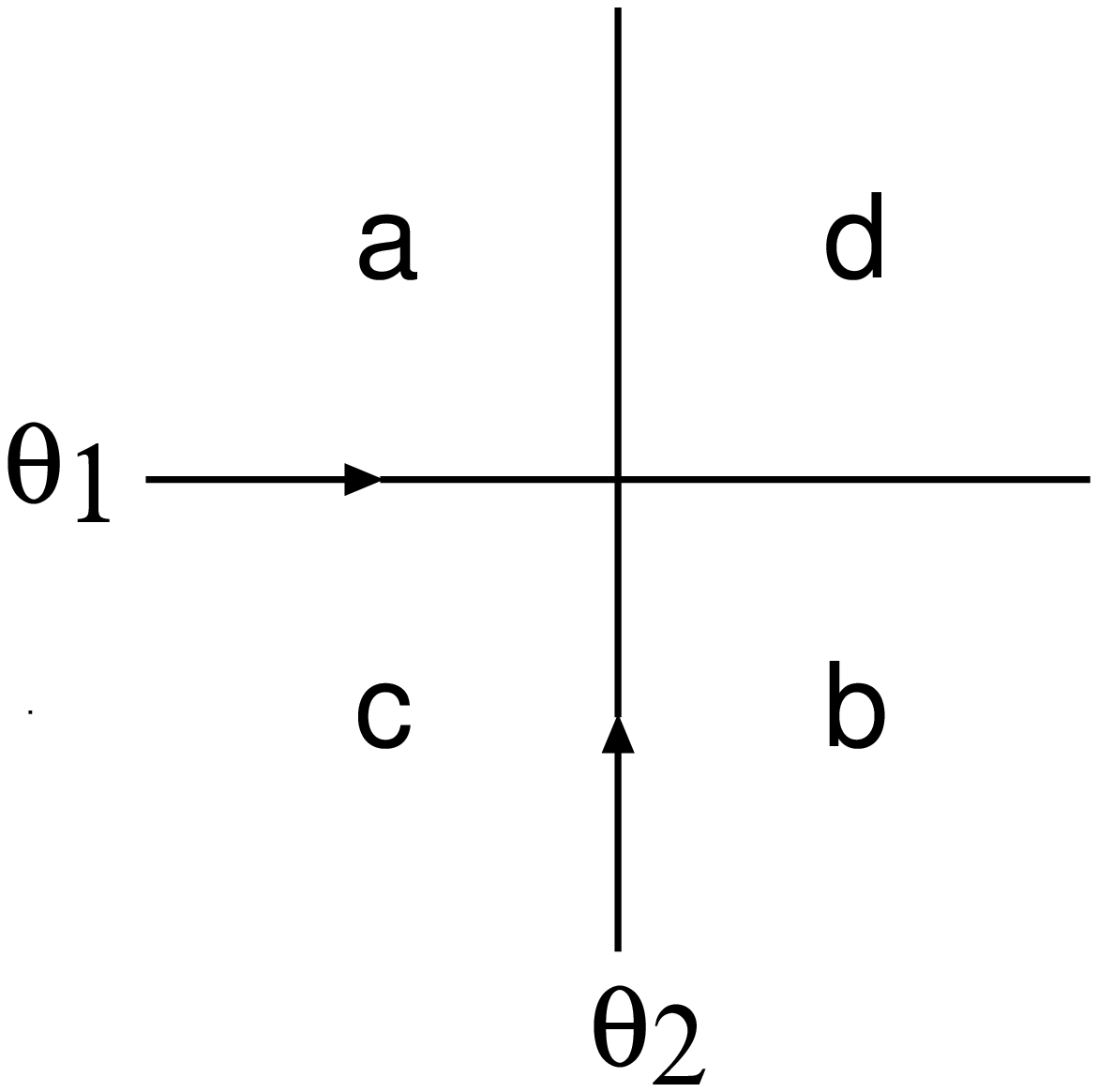}
	\caption[xxx]{\parbox[t]{0.4\textwidth}{
	Bulk $S$ matrix $S_{a\ b}^{c\ d}(\theta_{1}-\theta_{2})$.}
	}
	\label{fig1}
\end{figure}

The two-kink $S$ matrix $S_{a\ b}^{c\ d}(\theta)$
is defined by the relation (see Figure \ref{fig1})
\be
K_{a\,, c}(\theta_{1})\ K_{c\,, b}(\theta_{2}) =
\sum_{d} S_{a\ b}^{c\ d}(\theta_{1}-\theta_{2})\
K_{a\,, d}(\theta_{2})\ K_{d\,, b}(\theta_{1}).
\label{bulkS0}
\ee
The nonzero matrix elements are given by \cite{Za2, AlZa2} 
\footnote{It is noted in \cite{Za2} that this $S$ matrix is
essentially the solution of the star-triangle equation corresponding
to the critical Ising lattice model \cite{Ba, ABF}.}
\be
S_{0\ 0}^{\sigma\ \sigma'}(\theta) &=& e^{-i \gamma \theta} 
\sigma(\theta)\
\bar S_{0\ 0}^{\sigma\ \sigma'}(\theta) \,, \non \\
S_{\sigma\ \sigma'}^{0\ 0}(\theta) &=& e^{i \gamma \theta} 
\sigma(\theta)\
\bar S_{\sigma\ \sigma'}^{0\ 0}(\theta) \,,
\label{bulkS1}
\ee
where $\sigma \,, \sigma' \in \{ -1 \,, +1 \}$,  
$\gamma = {1\over 2 \pi}\ln 2$, and the ``reduced''
matrix elements $\bar S_{a\ b}^{c\ d}(\theta)$ are given by
\be
\bar S_{0\  0}^{\sigma\ \sigma}(\theta) &=&  \cosh {\theta\over 4} 
\,, \qquad \qquad \qquad 
\bar S_{0\ \ 0}^{\sigma\ -\sigma}(\theta) =  -i \sinh {\theta\over 4}
\,, \non \\
\bar S_{\sigma\ \sigma}^{0\ 0}(\theta) &=& 
\sqrt{2} \cosh {1\over 4}(\theta - i \pi) \,, \qquad 
\bar S_{\sigma\ -\sigma}^{0\ \ 0}(\theta) =
\sqrt{2} \cosh {1\over 4}(\theta + i \pi)
\,.
\label{bulkSreduced}
\ee
Finally, $\sigma(\theta)$ is a function which obeys 
\be
\sigma(\theta) = \sigma(i \pi - \theta) \,, \qquad
\sigma(\theta)\ \sigma(-\theta) = {1\over \cosh {\theta\over 2}} \,,
\label{sigmaproperties}
\ee
and has no poles in the physical strip $0 \le \Im m\ \theta < \pi$. A useful 
integral representation for this function is
\be
\sigma(\theta) = {-i\over \sqrt{2} \sinh({1\over 4}(\theta - i \pi))}
\exp \left( i \int_{0}^{\infty} {dt\over t} {\sin({\theta t\over \pi}) 
\sinh({3t\over 2})\over \sinh(2t) \cosh({t\over 2})} \right) \,.
\label{sigma}
\ee
This is a ``reduction'' of the well-known integral representation 
for the factor $U(\theta)$ of the sine-Gordon $S$ matrix \cite{ZZ}
with ${8 \pi\over \gamma'} = {1\over 4}$.

Zamolodchikov has shown in \cite{Za2} that this $S$ matrix ``commutes''
with supersymmetry charges $Q$ and $\bar Q$, which obey the $N=1$
supersymmetry algebra with topological charge.  The $S$ matrix also
commutes with the spin-reversal operator $\Gamma$, which is defined by
\be
\lefteqn{\Gamma\  K_{a_{1}\,, a_{2}}(\theta_{1})
\ K_{a_{2}\,, a_{3}}(\theta_{2}) 
\ldots K_{a_{N}\,, a_{N+1}}(\theta_{N})} \non \\
& & =
K_{-a_{1}\,, -a_{2}}(\theta_{1})\ K_{-a_{2}\,, -a_{3}}(\theta_{2}) 
\ldots K_{-a_{N}\,, -a_{N+1}}(\theta_{N})  \,.
\label{spinreversal}
\ee
Further properties of the $S$ matrix are listed in Appendix 
\ref{app:props}.

\subsection{Boundary}

Although the three vacua $-1 \,, 0 \,, +1$ are degenerate in the bulk,
these vacua do not necessarily remain degenerate at the boundary. 
Chim \cite{Ch} has identified the six conformal boundary conditions
(CBC) \cite{Ca} of the tricritical Ising CFT as follows: for the
boundary conditions $(-) \,, (0) \,, (+)$, the order parameter is
fixed at the boundary to the vacua $-1 \,, 0 \,, +1$, respectively. 
For the boundary condition $(-0)$, the vacua $-1$ and $0$ are
degenerate at the boundary; hence, the order parameter at the boundary
may be in either of these two vacua.  Similarly, for the boundary
condition $(0+)$, the $0$ and $+1$ vacua are degenerate at the
boundary.  Finally, for the boundary condition $(d)$, all three vacua
$-1 \,, 0 \,, +1$ are degenerate at the boundary (as well as in the
bulk); i.e., the order parameter at the boundary may be in any of the
three vacua.  The corresponding $g$ factors \cite{AL} are given by
\cite{Ch}
\be
g_{(d)} &=& \sqrt{2} \eta^{2} C \,, \qquad 
g_{(-0)} = g_{(0+)} =  \eta^{2} C \,,
\non \\
g_{(0)} &=&  \sqrt{2}  C \,, \qquad g_{(-)} = g_{(+)} =  C \,,
\label{gfactors}
\ee
where
\be
C = \sqrt{{\sin{\pi\over 5}}\over{\sqrt{5}}} \,, \qquad 
\eta = \sqrt{{\sin{{2\pi}\over 5}}\over{\sin{\pi\over 5}}} \,.
\ee
It is argued in \cite{Ne1} that the conformal boundary conditions $(-)
\& (+)$, $(-0) \& (0+)$, $(0)$ and $(d)$ are in fact superconformal.
Notice that the first two of these superconformal boundary conditions 
correspond to superpositions of ``pure'' Cardy states. 

\begin{figure}[tb]
	\centering
	\epsfxsize=0.20\textwidth\epsfbox{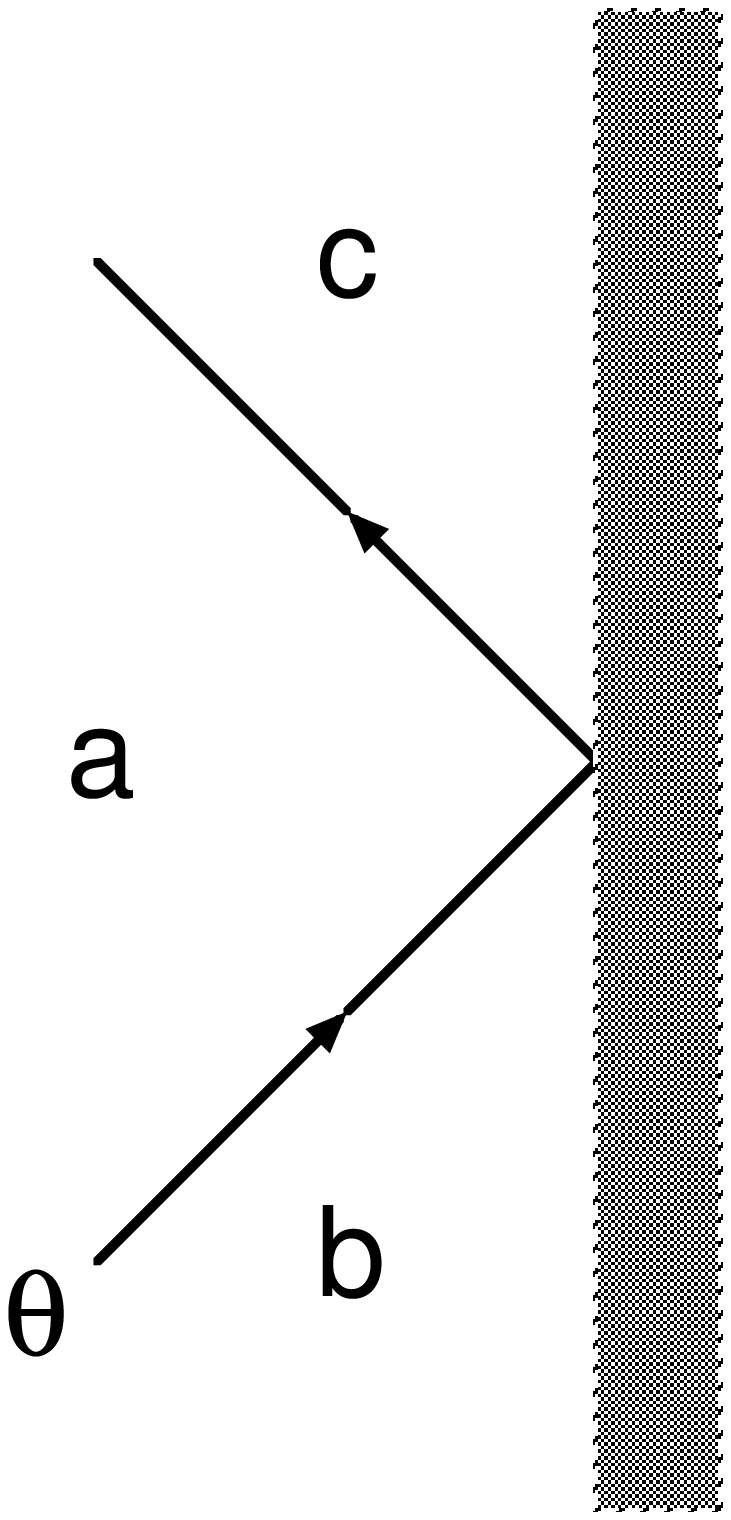}
	\caption[xxx]{\parbox[t]{0.4\textwidth}{
	Boundary $S$ matrix $\R{c}{a}{b}(\theta)$.}
	}
	\label{fig2}
\end{figure}

We shall consider separately integrable perturbations of both
conformal and superconformal boundary conditions, resulting in models
without and with supersymmetry, respectively.  We assume \cite{Ch}
that also in the perturbed theory the boundary can have (at most)
three possible states, corresponding to the three different vacua,
which are created by the boundary operator
$B_{a}$ with $a \in \{ -1 \,, 0 \,, 1\}$.  Multi-kink states have the
form
\be
K_{a_{1} \,, a_{2}}(\theta_{1})\ 
K_{a_{2} \,, a_{3}}(\theta_{2}) \ldots 
K_{a_{N} \,, a}(\theta_{N})\ B_{a} \,. \non 
\ee
The kink boundary $S$ matrix $\R{c}{a}{b}(\theta)$ is defined by
the relation (see Figure \ref{fig2})
\be
K_{a \,, b}(\theta)\ B_{b} = \sum_{c} \R{c}{a}{b}(\theta)\ 
K_{a\,, c}(-\theta)\ B_{c} \,.
\label{boundS0}
\ee

\subsubsection{Non-supersymmetric cases}\label{subsubsec:nosusy}

Chim \cite{Ch} has considered the TIM on the half-line $x \le 0$
corresponding to an integrable perturbation of the CBC $(-0)$.  The 
model is defined by the action
\be
A &=& A_{{\cal M}(4/5)+ (-0)} + \lambda \int_{-\infty}^{\infty} dy 
\int_{-\infty}^{0} dx\  
\Phi_{({3\over 5}\,, {3\over 5})}(x \,, y) \non \\
&-& h \int_{-\infty}^{\infty} dy\  
\phi_{({3\over 5})\,, (-0)}(y) \,, \qquad \lambda < 0 \,.
\label{boundaryaction1}
\ee
The last term is the boundary perturbation.  It involves the boundary
primary field $\phi_{({3\over 5})\,, (-0)}$ with dimension
$\Delta_{(1\,, 3)} = {3\over 5}$ which acts on the CBC $(-0)$. 
\footnote{In general, boundary operators $\phi_{a}$ and $\phi_{b}$
which act on conformal boundary conditions $a$ and $b$ commute; i.e.,
their operator product expansion with each other is zero.  Such
operators have recently been studied in \cite{Gr}.} Moreover, $h$ is a
boundary parameter which has dimensions length${}^{-{2\over 5}}$.

The boundary $S$ matrix which has been proposed \cite{Ch}
for this model has the following nonzero matrix elements
\be
\R{-1}{0}{-1}(\theta\,, \xi) &=& 
P(\theta\,, \xi)\ \barR{-1}{0}{-1}(\theta\,, \xi) \,, \non \\
\R{0}{\sigma}{0}(\theta\,, \xi) &=& 
M(\theta\,, \xi)\ \barR{0}{\sigma}{0}(\theta\,, \xi) \,,
\label{boundSnosusy1}
\ee
where the reduced matrix elements $\barR{c}{a}{b}(\theta\,, \xi)$
are given by
\be
\barR{-1}{0}{-1}(\theta\,, \xi) &=& 1 \,, \non \\
\barR{\ 0}{\pm 1}{\ 0}(\theta\,, \xi) &=& 
\cos {\xi\over 2} \pm i \sinh {\theta\over 2} \,.
\label{boundSnosusy2}
\ee
The parameter $\xi$ is related in some way to the boundary parameter 
$h$ appearing in the action (\ref{boundaryaction1}). The function 
$P(\theta\,, \xi)$ is given by
\be
P(\theta\,, \xi) = P^{CDD}(\theta \,, \xi)\  P_{min}(\theta) \,,
\label{Ptheta}
\ee
where $P^{CDD}(\theta\,, \xi)$ is the CDD factor
\be
P^{CDD}(\theta\,, \xi) = 
{\sin \xi - i \sinh \theta \over \sin \xi + i \sinh \theta} \,,
\label{Pcdd}
\ee
which has a pole at $\theta = i \xi$, and $P_{min}(\theta)$ is the
minimal solution of the equations
\be
P_{min}(\theta) P_{min}(-\theta) = 1 \,, \quad 
P_{min}({i \pi\over 2} - \theta) = 
\sqrt{2}e^{2i\gamma \theta} \cosh({\theta\over 2} - {i\pi\over 4}) 
\sigma(2 \theta) P_{min}({i \pi\over 2} + \theta) \,,
\label{Pminproperties}
\ee
with no poles in the physical strip $0 \le \Im m\ \theta < {\pi\over 2}$.
We find that it has the integral representation
\be
P_{min}(\theta) = \exp i\left( -\gamma \theta 
+ {1\over 8} \int_{0}^{\infty} {dt\over t} {\sin({2\theta t\over \pi}) 
\over \cosh^{2}t \cosh^{2} {t\over 2}} \right) \,.
\label{Pmin}
\ee
Finally, the function $M(\theta\,, \xi)$ is given by
\be
M(\theta\,, \xi) = e^{2i\gamma \theta}\sigma(\theta-i\xi)\ 
\sigma(\theta+i\xi)\ P(\theta\,, \xi) \,.
\label{Mtheta}
\ee

There is a similar model corresponding to a perturbation of the CBC
$(0+)$.  The boundary $S$ matrix for this case is the same as the one
given above, except that $\R{-1}{0}{-1}(\theta\,, \xi) =0$, and
\be
\R{1}{0}{1}(\theta\,, \xi) =
P(\theta\,, \xi)\ \barR{1}{0}{1}(\theta\,, \xi) \,, 
\label{otherboundSnosusy}
\ee 
with $\barR{1}{0}{1}(\theta\,, \xi) =1$.
Neither of these two models has supersymmetry.

\subsubsection{Supersymmetric cases}\label{subsubsec:susy}

Supersymmetric perturbations of the tricritical Ising boundary CFT
with two different superconformal boundary conditions (namely, $(-0)
\& (0+)$ and $(d)$) are considered in \cite{Ne1}.  We refer to these
two cases as NS and R, respectively, since these are the sectors to
which the corresponding boundary states belong.

\bigskip 

\noindent {\em NS case}

\medskip 

The NS case corresponds to a perturbation of the boundary condition
$(-0) \& (0+)$, with action
\be
A &=& A_{{\cal M}(4/5)+ (-0) \& (0+)} 
+ \lambda \int_{-\infty}^{\infty} dy 
\int_{-\infty}^{0} dx\  
\Phi_{({3\over 5}\,, {3\over 5})}(x \,, y) \non \\
&-& h \int_{-\infty}^{\infty} dy\  
\left( \phi_{({3\over 5})\,, (-0)}(y) 
- \phi_{({3\over 5})\,, (0+)}(y) \right) \,, \qquad \lambda < 0 \,.
\label{boundaryaction2}
\ee
The proposed boundary $S$ matrix is the ``direct sum'' of the boundary
$S$ matrices given in Sec.  \ref{subsubsec:nosusy} for the
perturbations of $(-0)$ and $(0+)$. That is, the nonzero matrix 
elements are given by
\be
\R{\sigma}{0}{\sigma}(\theta\,, \xi) &=& 
P(\theta\,, \xi)\ \barR{\sigma}{0}{\sigma}(\theta\,, \xi) 
\,, \non \\
\R{0}{\sigma}{0}(\theta\,, \xi) &=& 
M(\theta\,, \xi)\ \barR{0}{\sigma}{0}(\theta\,, \xi) \,,
\label{boundNS1}
\ee
where 
\be
\barR{\sigma}{0}{\sigma}(\theta\,, \xi) &=& 1 \,, \non \\
\barR{\ 0}{\pm 1}{\ 0}(\theta\,, \xi) &=& 
\cos {\xi\over 2} \pm i \sinh {\theta\over 2} \,,
\label{boundNS2}
\ee
and the functions $P(\theta\,, \xi)$ and $M(\theta\,, \xi)$ are
given by Eqs. (\ref{Ptheta}) and (\ref{Mtheta}), respectively.
This boundary $S$ matrix ``commutes'' with the supersymmetry charge
\be
\hat Q = Q + \bar Q + 2 \cos ({\xi\over 2}) \sqrt{m}\ \Gamma \,,
\label{hatQ1}
\ee
where $\Gamma$ is the spin-reversal operator (\ref{spinreversal}).

\bigskip 

\noindent {\em R case}

\medskip 

For the R case, which corresponds to a perturbation of the boundary
condition $(d)$, the action is given by the image of
(\ref{boundaryaction2}) under duality transformation.  The proposed
boundary $S$ matrix has the following nonzero matrix elements
\be
\R{\sigma}{0}{\sigma}(\theta\,, \xi) &=& 
N(\theta\,, \xi)\ \barR{\sigma}{0}{\sigma}(\theta\,, \xi) 
\,, \non \\
\R{-\sigma}{0}{\ \sigma}(\theta\,, \xi) &=& 
N(\theta\,, \xi)\ \barR{-\sigma}{0}{\ \sigma}(\theta\,, \xi) 
\,, \non \\
\R{0}{\sigma}{0}(\theta\,, \xi) &=& 
R(\theta\,, \xi)\ \barR{0}{\sigma}{0}(\theta\,, \xi) \,,
\label{boundR1}
\ee
where the reduced matrix elements $\barR{c}{a}{b}(\theta\,,
\xi)$ are given by
\be
\barR{\sigma}{0}{\sigma}(\theta\,, \xi) &=& \cos {\xi\over 2} 
\,, \non \\
\barR{-1}{0}{+1}(\theta\,, \xi) &=& -i r \sinh{\theta\over 2}
\,, \qquad 
\barR{+1}{0}{-1}(\theta\,, \xi) =-{i\over r} \sinh{\theta\over 2}
\,, \non \\
\barR{0}{\sigma}{0}(\theta\,, \xi) &=& 1 \,,
\label{boundR2}
\ee
and $r$ is a parameter which presumably is related in
some way to the boundary parameter $h$, as is $\xi$.  Moreover, the functions
$N(\theta\,, \xi)$ and $R(\theta\,, \xi)$ are given by
\be
N(\theta\,, \xi) &=& e^{-2i\gamma \theta} M(\theta\,, \xi) \,, \non \\
R(\theta\,, \xi) &=& e^{2i\gamma \theta} P(\theta\,, \xi) \,.
\label{NRtheta}
\ee
This boundary $S$ matrix ``commutes'' with the supersymmetry charge
\be
\hat Q = Q - \bar Q + {4 i r\over 1-r^{2}} \cos ({\xi\over 2}) 
\sqrt{m}\ \Gamma \,.
\label{hatQ2}
\ee
In contrast to the NS case, here the matrix $\R{c}{a}{b}(\theta\,,
\xi)$ does not vanish for $b \ne c$; i.e., it is not ``diagonal.''

The parameter $r$ can be set to unity by an appropriate gauge
transformation \cite{GZ} of the kink operators, which corresponds to
adding a total derivative term to the boundary action that restores
spin-reversal symmetry.  This limiting case, for which the
supersymmetry charge (\ref{hatQ2}) reduces to $\Gamma$, was considered
earlier in \cite{Ch}.

\section{Yang matrix and transfer matrix}\label{sec:yangtransfer}

The first step of the TBA program is to formulate the ``Yang matrix''
\cite{Ya} and relate it to an appropriate commuting transfer matrix. 
Since it is not obvious how to do this for the case of boundaries, we
begin by reviewing the case \cite{AlZa2} of periodic boundary
conditions. \footnote{The analysis presented here for RSOS-type $S$ 
matrices is parallel to the one given in \cite{AN} for vertex-type 
$S$ matrices.}

\subsection{Closed-chain transfer matrix}

Following \cite{AlZa2}, we consider $N$ kinks of mass $m$ with real
rapidities $\theta_{1} \,, \ldots \,, \theta_{N}$ and two-kink $S$
matrix $S_{a\ b}^{c\ d}(\theta)$ in a periodic box of
length $L >> {1\over m}$.  We impose the periodicity condition
\be
\lefteqn{e^{i L m \sinh \theta_{1}} K_{a_{1}\,, a_{2}}(\theta_{1}) 
K_{a_{2}\,, a_{3}}(\theta_{2}) \ldots K_{a_{N-1}\,, a_{N}}(\theta_{N-1}) 
K_{a_{N}\,, a_{1}}(\theta_{N})} \non \\
& & = K_{a_{2}\,, a_{3}}(\theta_{2}) \ldots
K_{a_{N}\,, a_{1}}(\theta_{N}) K_{a_{1}\,, a_{2}}(\theta_{1}) \,.
\label{closedassumption}
\ee
Commuting the kink operator $K_{a_{1}\,, a_{2}}(\theta_{1})$ on the
LHS past the others using the relation (\ref{bulkS0}), we obtain
\be
\lefteqn{e^{i L m \sinh \theta_{1}}
\sum_{d_{2}\,, \ldots \,, d_{N}} \Big\{ 
S_{a_{1}\ a_{3}}^{a_{2}\ d_{2}}(\theta_{1}-\theta_{2})
S_{d_{2}\ a_{4}}^{a_{3}\ d_{3}}(\theta_{1}-\theta_{3}) \ldots
S_{d_{N-2}\ a_{N}}^{a_{N-1}\ d_{N-1}}(\theta_{1}-\theta_{N-1}) 
S_{d_{N-1}\ a_{1}}^{a_{N}\ \ d_{N}}(\theta_{1}-\theta_{N})}\non \\
& & K_{a_{1}\,, d_{2}}(\theta_{2}) 
K_{d_{2}\,, d_{3}}(\theta_{3}) \ldots
K_{a_{N-1}\,, d_{N}}(\theta_{N}) K_{d_{N}\,, a_{1}}(\theta_{1}) 
\Big\} = K_{a_{2}\,, a_{3}}(\theta_{2}) \ldots
K_{a_{N}\,, a_{1}}(\theta_{N}) K_{a_{1}\,, a_{2}}(\theta_{1}) \,.
\non \\
\ee
Multiplying both sides by the ``wavefunction'' $\Psi^{a_{1} \ldots 
a_{N}}$, summing over $a_{1} \,, \ldots \,, a_{N}$, and relabeling 
indices appropriately, we obtain the Yang equation for kink 1
\be
e^{i L m \sinh \theta_{1}}
\sum_{a'_{1}\,,  \ldots \,, a'_{N}} 
Y_{(1)\ a'_{1} \ldots a'_{N}}^{\ \ \ a_{1} \ldots a_{N}}\ 
\Psi^{a'_{1} \ldots a'_{N}} = \Psi^{a_{1} \ldots a_{N}} \,,
\ee
where $Y_{(1)}$ is the Yang matrix
\be
Y_{(1)\ a'_{1} \ldots a'_{N}}^{\ \ \ a_{1} \ldots a_{N}} = 
\delta^{a_{2}}_{a'_{1}}
S_{a_{2}\ a'_{3}}^{a'_{2}\ a_{3}}(\theta_{1}-\theta_{2})
S_{a_{3}\ a'_{4}}^{a'_{3}\ a_{4}}(\theta_{1}-\theta_{3}) \ldots
S_{a_{N-1}\ a'_{N}}^{a'_{N-1}\ a_{N}}(\theta_{1}-\theta_{N-1}) 
S_{a_{N}\ a_{2}}^{a'_{N}\ a_{1}}(\theta_{1}-\theta_{N}) \,.
\ee
There are similar equations, and 
corresponding matrices $Y_{(k)}$, for the other kinks $k = 2 \,, 
3 \,, \ldots \,, N$.

The objective is to diagonalize $Y_{(k)}$.  The key to this problem is
to relate $Y_{(k)}$ to an inhomogeneous closed-chain transfer matrix,
for which there are well-developed diagonalization techniques.  To
this end, we consider the transfer matrix
\be
\tau_{a'_{1} \ldots a'_{N}}^{a_{1} \ldots a_{N}}
(\theta | \theta_{1} \,, \ldots \,, \theta_{N})
= 
S_{a_{1}\ a'_{2}}^{a'_{1}\ a_{2}}(\theta-\theta_{1})
S_{a_{2}\ a'_{3}}^{a'_{2}\ a_{3}}(\theta-\theta_{2}) \ldots
S_{a_{N-1}\ a'_{N}}^{a'_{N-1}\ a_{N}}(\theta-\theta_{N-1}) 
S_{a_{N}\ a'_{1}}^{a'_{N}\ a_{1}}(\theta-\theta_{N}) \,,
\label{closedtransfer}
\ee
with inhomogeneities $\theta_{1} \,, \ldots \,, \theta_{N}$.  Because
the $S$ matrix satisfies the Yang-Baxter equation (\ref{YBE}), the
transfer matrix commutes for different values of $\theta$ 
\footnote{Our convention for matrix multiplication is given by
\be
(A B)_{a'_{1} \ldots a'_{N}}^{a_{1} \ldots a_{N}} =
\sum_{a''_{1}\,,  \ldots \,, a''_{N}} 
A_{a'_{1} \ldots a'_{N}}^{a''_{1} \ldots a''_{N}}\
B_{a''_{1} \ldots a''_{N}}^{a_{1} \ldots a_{N}} \non \,.
\ee}
\be
\left[ \tau(\theta | \theta_{1} \,, \ldots \,, \theta_{N}) \,,
\tau(\theta' | \theta_{1} \,, \ldots \,, \theta_{N})  
\right] = 0 \,.
\label{closedcommutativity}
\ee

Let us now evaluate this transfer matrix at $\theta = \theta_{1}$. 
Using the fact that the $S$ matrix at zero rapidity is given by 
(\ref{zerorapid}), we immediately obtain 
$\tau(\theta_{1} | \theta_{1} \,, \ldots \,, \theta_{N}) = 
Y_{(1)}$. In general, we have
\be
Y_{(k)} = \tau(\theta_{k} | \theta_{1} \,, \ldots \,, \theta_{N})
\,, \qquad k = 1 \,,  \ldots \,, N \,.
\label{closedresult}
\ee
This is the sought-after relation. In order to diagonalize the Yang 
matrices $Y_{(k)}$, it suffices to diagonalize the commuting 
closed-chain transfer matrix 
$\tau(\theta | \theta_{1} \,, \ldots \,, \theta_{N})$.
That calculation, as well as the corresponding bulk TBA analysis, is 
described for the TIM in \cite{AlZa2}.

\subsection{Open-chain transfer matrix}

We turn now to the case with boundaries, which is our primary interest
here.  We therefore consider $N$ kinks of mass $m$ with
real, positive rapidities $\theta_{1} \,, \ldots \,, \theta_{N}$ in an interval
of length $L >> {1\over m}$, with bulk $S$ matrix $S_{a\ b}^{c\ d}(\theta)$
and boundary $S$ matrix $\R{c}{a}{b}(\theta\,, \xi)$.  In analogy 
with (\ref{closedassumption}), we propose the formal relation
\be
\lefteqn{e^{2 i L m \sinh \theta_{1}} 
B_{a_{1}}^{+}\ K_{a_{1}\,, a_{2}}(\theta_{1}) 
K_{a_{2}\,, a_{3}}(\theta_{2}) \ldots K_{a_{N-1}\,, a_{N}}(\theta_{N-1}) 
K_{a_{N}\,, a_{N+1}}(\theta_{N})\ B_{a_{N+1}}^{-}} \non \\
& & = B_{a_{1}}^{+}\ K_{a_{1}\,, a_{2}}(\theta_{1}) 
K_{a_{2}\,, a_{3}}(\theta_{2}) \ldots K_{a_{N-1}\,, a_{N}}(\theta_{N-1}) 
K_{a_{N}\,, a_{N+1}}(\theta_{N})\ B_{a_{N+1}}^{-} \,,
\label{openassumption}
\ee
where now there are two boundary operators $B_{a}^{\pm}$ corresponding 
to the left and right boundaries, with (cf, Eq. (\ref{boundS0})) 
\footnote{The relations (\ref{boundSminus}) and (\ref{boundSplus}) are 
consistent in that both lead to the same boundary Yang-Baxter 
equation (\ref{BYBE}).}
\be
K_{a \,, b}(\theta)\ B_{b}^{-} &=& 
\sum_{c} \R{c}{a}{b}(\theta \,, \xi_{-})\ 
K_{a\,, c}(-\theta)\ B_{c}^{-} \,,  \label{boundSminus} \\
B_{b}^{+}\ K_{b \,, a}(\theta)  &=& 
\sum_{c} B_{c}^{+}\ K_{c\,, a}(-\theta)\ 
\R{c}{a}{b}(-\theta \,, \xi_{+}) \,.  \label{boundSplus}
\ee
Note that for each boundary operator $B_{a}^{\pm}$ there is a
corresponding boundary parameter $\xi_{\pm}$.
By moving the kink operator with rapidity $\theta_{1}$ on the LHS of
(\ref{openassumption}) to the far right using (\ref{bulkS0}),
reflecting it from the right boundary using (\ref{boundSminus}),
moving it to the far left using again (\ref{bulkS0}), and finally
reflecting it from the left boundary using (\ref{boundSplus}), we
arrive at the Yang equation for kink 1
\be
e^{2 i L m \sinh \theta_{1}}
\sum_{a'_{1}\,,  \ldots \,, a'_{N+1}} 
Y_{(1)\ a'_{1} \ldots a'_{N+1}}^{\ \ \ a_{1} \ldots a_{N+1}}\ 
\Psi^{a'_{1} \ldots a'_{N+1}} = \Psi^{a_{1} \ldots a_{N+1}} \,,
\label{openYang1}
\ee
where the Yang matrix $Y_{(1)}$ is given by
\be
Y_{(1)\ a'_{1} \ldots a'_{N+1}}^{\ \ \ a_{1} \ldots a_{N+1}}
&=& \sum_{d_{2} \,, \ldots \,, d_{N}} \Big\{
\R{a_{1}}{a_{2}}{a'_{1}}(\theta_{1} \,, \xi_{+})\ 
S_{a'_{1}\ a'_{3}}^{a'_{2}\ d_{2}}(\theta_{1}-\theta_{2}) \ldots
S_{d_{N-1}\ a'_{N+1}}^{a'_{N}\ \ \ d_{N}}(\theta_{1}-\theta_{N}) \non \\
&\times& \R{a_{N+1}}{d_{N}}{a'_{N+1}}(\theta_{1} \,, \xi_{-})\
S_{d_{N-1}\ a_{N+1}}^{d_{N}\ \ \ a_{N}}(\theta_{1}+\theta_{N}) \ldots
S_{a'_{1}\ a_{3}}^{d_{2}\ a_{2}}(\theta_{1}+\theta_{2}) \Big\} \,.
\ee
There are similar matrices $Y_{(k)}$ for the other kinks. In 
analogy with the case of periodic boundary conditions, the key to 
diagonalizing the Yang matrix is to relate it to an inhomogeneous 
open-chain transfer matrix 
\be
t_{a'_{1} \ldots a'_{N+1}}^{a_{1} \ldots a_{N+1}}
(\theta | \theta_{1} \,, \ldots \,, \theta_{N})
&=& \sum_{a''_{1}\,,  \ldots \,, a''_{N+1}} \Big\{
\R{a_{1}}{a''_{1}}{a'_{1}}(i \pi -\theta \,, \xi_{+})\ 
S_{a''_{1}\ a'_{2}}^{a'_{1}\ a''_{2}}(\theta-\theta_{1}) \ldots
S_{a''_{N}\ a'_{N+1}}^{a'_{N}\ a''_{N+1}}(\theta-\theta_{N}) \non \\
&\times& \R{\ a_{N+1}}{a''_{N+1}}{\ a'_{N+1}}(\theta \,, \xi_{-})\
S_{a''_{N}\ a_{N+1}}^{a''_{N+1}\ a_{N}}(\theta +\theta_{N}) \ldots
S_{a''_{1}\ a_{2}}^{a''_{2}\ a_{1}}(\theta+\theta_{1}) \Big\} \,,
\label{transfer}
\ee
which commutes for different values of $\theta$
\be
\left[ t(\theta | \theta_{1} \,, \ldots \,, \theta_{N}) \,,
t(\theta' | \theta_{1} \,, \ldots \,, \theta_{N})  
\right] = 0 \,.
\label{opencommutativity}
\ee
The transfer matrix (\ref{transfer}) is an RSOS version \cite{AK, BPO, 
Zh} of the Sklyanin \cite{Sk} vertex-type transfer matrix. 
Using the relations (\ref{zerorapid}), (\ref{boundcrossunit}) and
(\ref{crossing}), one can show that
\be
Y_{(k)} = t(\theta_{k} | \theta_{1} \,, \ldots \,, \theta_{N})
\,, \qquad k = 1 \,,  \ldots \,, N \,.
\label{openresult}
\ee
Hence, in order to diagonalize the Yang matrices
$Y_{(k)}$, it suffices to diagonalize the open-chain 
transfer matrix $t(\theta | \theta_{1} \,, \ldots \,, \theta_{N})$.
Indeed, let $\Psi(\theta_{1} \,, \ldots \,, \theta_{N})$ be
an eigenvector of the transfer matrix with corresponding eigenvalue
$\Lambda(\theta | \theta_{1} \,, \ldots \,, \theta_{N})$,
\be
t(\theta | \theta_{1} \,, \ldots \,, \theta_{N})\  
\Psi(\theta_{1} \,, \ldots \,, \theta_{N}) =
\Lambda(\theta | \theta_{1} \,, \ldots \,, \theta_{N})\ 
\Psi(\theta_{1} \,, \ldots \,, \theta_{N}) \,.
\label{eigenvector}
\ee
The eigenvector is independent of $\theta$ by virtue of the
commutativity property (\ref{opencommutativity}).  With the help of
the result (\ref{openresult}), the Yang equation (\ref{openYang1})
implies
\be
e^{2 i L m \sinh \theta_{k}}
\Lambda(\theta_{k} | \theta_{1} \,, \ldots \,, \theta_{N}) = 1 \,,
\qquad k = 1 \,,  \ldots \,, N \,.
\label{openYang2}
\ee

\section{Inversion identity and transfer-matrix 
eigenvalues: NS case}\label{sec:inversion}

We turn now to the problem of determining the eigenvalues of the
inhomogeneous open-chain transfer matrix (\ref{transfer}).  As for the
closed chain \cite{AlZa2}, our approach is to derive an exact
inversion identity.  For definiteness, we treat here the NS case. 
(See Sec.  \ref{subsubsec:susy}.)  The results for the R case, which
are closely related to those for the NS case, are presented in Sec. 
\ref{sec:Rcase}.

Instead of working with the full (``dressed'') transfer matrix
(\ref{transfer}), it is convenient (see Footnote 
\ref{period} below) to work instead with the reduced
(``bare'') transfer matrix $\bar t$, which is constructed from the
reduced bulk and boundary $S$ matrices,
\be
\bar t_{a'_{1} \ldots a'_{N+1}}^{a_{1} \ldots a_{N+1}}
(\theta | \theta_{1} \,, \ldots \,, \theta_{N})
&=& \sum_{a''_{1}\,,  \ldots \,, a''_{N+1}} \Big\{
\barR{a_{1}}{a''_{1}}{a'_{1}}(i \pi -\theta \,, \xi_{+})\ 
\bar S_{a''_{1}\ a'_{2}}^{a'_{1}\ a''_{2}}(\theta-\theta_{1}) \ldots
\bar S_{a''_{N}\ a'_{N+1}}^{a'_{N}\ a''_{N+1}}(\theta-\theta_{N}) \non \\
&\times& \barR{\ a_{N+1}}{a''_{N+1}}{\ a'_{N+1}}(\theta \,, \xi_{-})\
\bar S_{a''_{N}\ a_{N+1}}^{a''_{N+1}\ a_{N}}(\theta +\theta_{N}) \ldots
\bar S_{a''_{1}\ a_{2}}^{a''_{2}\ a_{1}}(\theta+\theta_{1}) \Big\} \,.
\label{reducedtransfer}
\ee
It is also convenient to define the following four ``sectors'':
\be
N=\mbox{ even } - \mbox{ Sector } I &:& 
a_{1}\,, a'_{1}\,, a_{N+1}\,, a'_{N+1}
\in \{-1 \,, +1\} \,, \non  \\
N=\mbox{ even } - \mbox{ Sector } II &:& 
a_{1} = a'_{1} = a_{N+1} = a'_{N+1} = 0 
\,, \non  \\
N=\mbox{ odd } - \mbox{ Sector } I &:& 
a_{1}\,, a'_{1} \in \{-1 \,, +1\}\,, 
\qquad a_{N+1} = a'_{N+1} = 0 \,, \non  \\
N=\mbox{ odd } - \mbox{ Sector } II &:& a_{1} = a'_{1} = 0 \,, 
\qquad a_{N+1} \,, a'_{N+1} \in \{-1 \,, +1\}  \,.
\label{sectors}
\ee 
The nonzero matrix elements of the transfer matrix lie exclusively in
these sectors.  For a given parity of $N$ (i.e., even or odd), the
transfer matrix decomposes into two blocks along the diagonal
corresponding to sectors I and II.
For the NS case (\ref{boundNS1}), (\ref{boundNS2}), the relation
between the full transfer matrix and the reduced transfer matrix is
given by
\be
t^{(\alpha)}(\theta | \theta_{1} \,, \ldots \,, \theta_{N}) =
w^{(\alpha)}(\theta)\ \bar t^{(\alpha)}
(\theta | \theta_{1} \,, \ldots \,, \theta_{N})
\,, \label{fullreducedreltn1}
\ee
where $\alpha$ runs over the four sectors (\ref{sectors}), and 
$w^{(\alpha)}(\theta)$ is given by
\be
w^{(\alpha)}(\theta) =
\prod_{j=1}^{N}\sigma(\theta-\theta_{j}) \sigma(\theta+\theta_{j}) 
\times
\left\{ \begin{array}{l}
          P(i\pi-\theta \,, \xi_{+})\ P(\theta \,, \xi_{-}) \\
	  M(i\pi-\theta \,, \xi_{+})\ M(\theta \,, \xi_{-})\\
	  e^{-2i \gamma \theta}P(i\pi-\theta \,, \xi_{+})\ 
	  M(\theta \,, \xi_{-}) \\
	  e^{2i \gamma \theta} M(i\pi-\theta \,, \xi_{+})\ 
	  P(\theta \,, \xi_{-}) \end{array}\right. \,,
\ee
respectively. The latter can be brought to the form 
\be
w^{(\alpha)}(\theta) = {1\over \sigma(2\theta) 
\cosh{\theta\over 2}} 
\prod_{j=1}^{N}\sigma(\theta-\theta_{j}) \sigma(\theta+\theta_{j}) 
\times
\left\{ \begin{array}{l}
          e^{2i \gamma \theta} P(\theta \,, \xi_{+})\ P(\theta \,, \xi_{-}) \\
	  {1\over 2}e^{-2i \gamma \theta} M(\theta \,, \xi_{+})\ M(\theta \,, \xi_{-})\\
	 P(\theta \,, \xi_{+})\ M(\theta \,, \xi_{-})\\
	   {1\over 2} M(\theta \,, \xi_{+})\ P(\theta \,, \xi_{-})
	   \end{array}\right.  \label{fullreducedreltn2}
\ee
with the help of the crossing properties (\ref{sigmaproperties}), 
(\ref{Pminproperties}).

Using the fusion procedure, we show  in Appendix \ref{app:fusion}
that the reduced transfer matrix obeys the inversion identity
\be
\bar t^{(\alpha)}(\theta | \theta_{1} \,, \ldots \,, \theta_{N})\
\bar t^{(\alpha)}(\theta + i\pi | \theta_{1} \,, \ldots \,, \theta_{N})
= f^{(\alpha)}(\theta)\ \id^{(\alpha)} \,,
\label{inversionid1}
\ee 
where $\alpha$ runs over the four sectors (\ref{sectors}), and 
$f^{(\alpha)}(\theta)$ is given by
\be
f^{(\alpha)}(\theta) &=& {1\over \cosh \theta}\Big[
f^{(\alpha)}_{+}(\theta) \cosh^{2}{\theta\over 2}
\prod_{j=1}^{N}\cosh({1\over 2}(\theta-\theta_{j}))
\cosh({1\over 2}(\theta+\theta_{j})) \non \\
&+& f^{(\alpha)}_{-}(\theta) \sinh^{2}{\theta\over 2}
\prod_{j=1}^{N}\sinh({1\over 2}(\theta-\theta_{j}))
\sinh({1\over 2}(\theta+\theta_{j})) \Big] \,,
\label{inversionid2}
\ee
where
\be
f^{(\alpha)}_{\pm}(\theta) =
\left\{ \begin{array}{c}
         1 \\
	 (\cosh \theta \pm \cos \xi_{-}) (\cosh \theta \pm \cos \xi_{+})\\
	{1\over 2} (\cosh \theta \pm \cos \xi_{-})\\
	 2 (\cosh \theta \pm \cos \xi_{+})
	   \end{array}\right. \,, \label{inversionid3}
\ee
respectively.  This inversion identity is one of the main results of
this paper.  We have checked it explicitly up to $N=4$.

In addition to the inversion identity, we can establish certain
further properties of the transfer matrix which are needed to
determine its eigenvalues.  Namely, periodicity \footnote{\label{period}
This is not the case for the full transfer matrix $t(\theta |
\theta_{1} \,, \ldots \,, \theta_{N})$.}
\be
\bar t(\theta + 2 i\pi| \theta_{1} \,, \ldots \,, \theta_{N}) = 
\bar t(\theta | \theta_{1} \,, \ldots \,, \theta_{N}) \,,
\label{periodicity}
\ee
crossing
\be
\bar t( i\pi - \theta | \theta_{1} \,, \ldots \,, \theta_{N}) = 
\bar t(\theta | \theta_{1} \,, \ldots \,, \theta_{N}) \,,
\label{transfercrossing}
\ee
and asymptotic behavior for large $\theta$
\be
\bar t^{(\alpha)} (\theta | \theta_{1} \,, \ldots \,, \theta_{N}) 
\sim z^{(\alpha)}(\theta)\ \id^{(\alpha)} \quad \mbox{ for } 
\quad \theta \rightarrow \infty \,, \label{asymptotic1}
\ee
where $\alpha$ runs over the four sectors (\ref{sectors}), and $z^{(\alpha)}$
is given by
\be
z^{(\alpha)}(\theta) = \left\{ \begin{array}{l}
          \left(-{i e^{\theta}\over 4}\right)^{N\over 2}
	  (\delta_{a_{1}\,, a_{N+1}}-\delta_{a_{1}\,, -a_{N+1}})\\
	   2\left(-{i e^{\theta}\over 4}\right)^{{N\over 2}+1}\\
	  \left(-{i e^{\theta}\over 4}\right)^{N+1\over 2}
	  (\delta_{a_{1}\,, -1}-\delta_{a_{1}\,, 1})\\
           2\left(-{i e^{\theta}\over 4}\right)^{N+1\over 2}
	  (\delta_{a_{N+1}\,, -1}-\delta_{a_{N+1}\,, 1})   \end{array}\right. 
\,, \label{asymptotic2}
\ee 
respectively.

Acting with the above relations on an eigenvector $\Psi(\theta_{1} \,, 
\ldots \,, \theta_{N})$ of the (reduced) transfer 
matrix
\be 
\bar t(\theta | \theta_{1} \,, \ldots \,, \theta_{N})\  
\Psi(\theta_{1} \,, \ldots \,, \theta_{N}) =
\bar \Lambda(\theta | \theta_{1} \,, \ldots \,, \theta_{N})\ 
\Psi(\theta_{1} \,, \ldots \,, \theta_{N}) \,,
\ee
we obtain corresponding relations for the eigenvalues
$\bar \Lambda^{(\alpha)}(\theta | \theta_{1} \,, \ldots \,, \theta_{N})$
in the various sectors,
\be
\bar \Lambda^{(\alpha)}(\theta | \theta_{1} \,, \ldots \,, \theta_{N})\
\bar \Lambda^{(\alpha)}(\theta + i\pi | \theta_{1} \,, \ldots \,, \theta_{N})
&=& f^{(\alpha)}(\theta) \,, \label{inversioneigen} \\
\bar \Lambda^{(\alpha)}(\theta + 2 i\pi| \theta_{1} \,, \ldots \,, \theta_{N}) 
&=& 
\bar \Lambda^{(\alpha)}(\theta | \theta_{1} \,, \ldots \,, \theta_{N}) \,, 
\label{periodicityeigen}  \\
\bar \Lambda^{(\alpha)}( i\pi - \theta | \theta_{1} \,, \ldots \,, \theta_{N}) 
&=&  
\bar \Lambda^{(\alpha)}(\theta | \theta_{1} \,, \ldots \,, \theta_{N}) 
\,, \label{crossingeigen}  \\
\bar \Lambda^{(\alpha)} (\theta | \theta_{1} \,, \ldots \,, \theta_{N}) 
&\sim& z^{(\alpha)}(\theta) \quad \mbox{ for } 
\quad \theta \rightarrow \infty \,. \label{asymptoticeigen}
\ee

The periodicity, crossing and asymptotic behavior requirements of the eigenvalues
(\ref{periodicityeigen}) - (\ref{asymptoticeigen}) are fulfilled by the Ansatz
\be
\bar \Lambda^{(\alpha)}(\theta | \theta_{1} \,, \ldots \,, \theta_{N}) 
= c^{(\alpha)} \prod_{j=1}^{d^{(\alpha)}} (-i) \sinh({1\over 2}(\theta - u_{j}))
\cosh({1\over 2}(\theta + u_{j})) \,,
\label{Ansatz}
\ee 
where $c^{(\alpha)}$ and $d^{(\alpha)}$ are given by
\be
c^{(\alpha)} = \left\{ \begin{array}{c}
          \pm 1\\
	   2\\
	  \pm 1\\
          \pm 2\end{array}\right. 
\,, \qquad 
d^{(\alpha)} = \left\{ \begin{array}{c}
          {N\over 2}\\
	   {N\over 2}+ 1\\
	  {N+ 1\over 2}\\
           {N+ 1\over 2}\end{array}\right. 
\,, \label{calphadalpha}
\ee 
respectively.
The parameters $\{ u_{j} \}$ appearing in the Ansatz (\ref{Ansatz}) are 
evidently roots of the eigenvalues, 
$\bar \Lambda^{(\alpha)}(u_{j} | \theta_{1} \,, \ldots \,, 
\theta_{N})=0$. It follows from the inversion identity (\ref{inversioneigen})
that $\{ u_{j} \}$ are also roots of the function $f^{(\alpha)}(\theta)$, 
i.e.,  $f^{(\alpha)}(u_{j})=0$. We conclude from (\ref{inversionid2}) that 
$\{ u_{j} \}$ are solutions of the set of equations
\be
-{f^{(\alpha)}_{-}(u_{j})\over f^{(\alpha)}_{+}(u_{j})}
{\sinh^{2}{u_{j}\over 2}\over  \cosh^{2}{u_{j}\over 2}}
\prod_{k=1}^{N}
{\sinh({1\over 2}(u_{j} -\theta_{k}))\over
\cosh({1\over 2}(u_{j} -\theta_{k}))}
{\sinh({1\over 2}(u_{j} +\theta_{k}))\over
\cosh({1\over 2}(u_{j} +\theta_{k}))} = 1 \,, \qquad j = 1\,, \ldots 
\,, d^{(\alpha)} \,, 
\label{BA1}
\ee
to which we refer as ``Bethe Ansatz'' equations.

The periodicity property (\ref{periodicityeigen}) implies that we can
restrict the roots $u_{j}$ of $\bar \Lambda^{(\alpha)}(\theta)$ to the 
interval
\be
-\pi < \Im m\ u_{j} \le \pi \,.
\ee
We now demonstrate that all the roots $u_{j}$ have the form $x_{j} \pm {i 
\pi\over 2}$ with $x_{j}$ real. Indeed, we observe that 
$f^{(\alpha)}(\theta)$ has the properties \footnote{We assume here that 
$\{ \theta_{k} \}$ and $\xi_{\pm}$ are real.}
\be
\left[ f^{(\alpha)}(\theta ) \right]^{*} = f^{(\alpha)}(\theta^{*}) 
\,, \qquad 
f^{(\alpha)}(\theta \mp i\pi) = f^{(\alpha)}(\theta ) \,,
\ee
where ${}^{*}$ denotes complex conjugation. These two properties 
imply that if $u_{j}$ is a root of $f^{(\alpha)}(\theta)$, then so 
are $u_{j}^{*}$ and $u_{j} \mp i\pi$, respectively. Since 
$u_{j}^{*} \ne u_{j}$, then $u_{j}^{*} = u_{j} \mp i\pi$. Hence, 
$\Im m\ u_{j} = \pm {i \pi\over 2}$.

In view of the above, we set
\be
u_{j} = x_{j} + {i\pi\over 2} \epsilon_{j}  \,,
\ee
with $x_{j}$ real and $\epsilon_{j} = \pm 1$. The eigenvalues 
(\ref{Ansatz}) are specified by 
$\{ x_{j} \,,  \epsilon_{j} \} \,, \quad j = 1 \,, \ldots \,, d^{(\alpha)}$,
similarly to the bulk case \cite{AlZa2}. Hence, we can rewrite the 
expression for the eigenvalues as
\be
\bar \Lambda^{(\alpha)}(\theta | \theta_{1} \,, \ldots \,, \theta_{N}) 
= c^{(\alpha)} \prod_{j=1}^{d^{(\alpha)}} 
\lambda^{(1)}_{\epsilon_{j}}(\theta - x_{j})\ 
\lambda^{(2)}_{\epsilon_{j}}(\theta + x_{j})\,,
\label{Ansatz2}
\ee 
where
\be
\lambda^{(1)}_{\epsilon}(\theta) = 
\sinh {1\over 2}(\theta - {i\pi\over 2} \epsilon) \,, \qquad
\lambda^{(2)}_{\epsilon}(\theta) = 
-i\cosh {1\over 2}(\theta + {i\pi\over 2} \epsilon) \,.
\label{lambdas}
\ee 
Moreover, we can rewrite the Bethe Ansatz 
equations (\ref{BA1}) in terms of $x_{j}$ (they do not depend
on $\epsilon_{j}$) as
\be
{\cal Q}^{(\alpha)}(x_{j} \,, \xi_{\pm})
{\sinh^{2}({x_{j}\over 2} - {i \pi\over 4})\over  
\sinh^{2}({x_{j}\over 2} + {i \pi\over 4})}
\prod_{k=1}^{N}\left[
-{\sinh({x_{j} -\theta_{k}\over 2}- {i \pi\over 4})\over
\sinh({x_{j} -\theta_{k}\over 2}+ {i \pi\over 4})}
{\sinh({x_{j} +\theta_{k}\over 2}- {i \pi\over 4})\over
\sinh({x_{j} +\theta_{k}\over 2}+ {i \pi\over 4})} \right]
= 1 \,, \non \\  
j = 1\,, \ldots  \,, d^{(\alpha)} \,, 
\label{BA2}
\ee
where 
\be
{\cal Q}^{(\alpha)}(x_{j} \,, \xi_{\pm}) =
 \left\{ \begin{array}{c}
          1\\
	  Q(x_{j} \,, \xi_{-})\  Q(x_{j} \,, \xi_{+})\\
	  Q(x_{j} \,, \xi_{-})\\
          Q(x_{j} \,, \xi_{+})\end{array}\right. 
\,, \label{calQ}
\ee
and
\be
Q(x \,, \xi) ={\sinh x - i \cos \xi\over \sinh x + i \cos \xi} \,.
\label{Qx}
\ee
(As always,  $\alpha$ runs over the four sectors (\ref{sectors}).) 
Notice that (\ref{BA2}) is invariant under $x_{j} \mapsto -x_{j}$. 
Moreover, following \cite{FS, GMN}, we assume that the root $x_{j}=0$
corresponds to an eigenvector with zero norm. Hence, we restrict to 
solutions with $x_{j} > 0$.

To summarize this Section, the eigenvalues (\ref{eigenvector}) of the
full transfer matrix for the NS case of the TIM are given by
\be
\Lambda^{(\alpha)}(\theta | \theta_{1} \,, \ldots \,, \theta_{N}) =
w^{(\alpha)}(\theta)\
\bar \Lambda^{(\alpha)}(\theta | \theta_{1} \,, \ldots \,, \theta_{N}) 
\label{eigenvaluefinal}
\,,
\ee
where $w^{(\alpha)}(\theta)$ is given by (\ref{fullreducedreltn2}), 
$\bar \Lambda^{(\alpha)}(\theta | \theta_{1} \,, \ldots \,, \theta_{N})$
is given by (\ref{Ansatz2}), and $\{ x_{j} \}$ are positive solutions of the 
Bethe Ansatz equations (\ref{BA2}).

\section{TBA analysis}\label{sec:TBA}

Having obtained the eigenvalues of the transfer matrix and the Bethe
Ansatz equations, we can proceed to the derivation of the
corresponding TBA equations and boundary entropy.  Following
\cite{GZ, LMSS} we consider the partition function $Z_{+-}$ of
the system on a cylinder of length $L$ and circumference $R$ with
left/right boundary conditions denoted by $\pm$.  It is given by
\be
Z_{+-} &=& \tr e^{-R H_{+-}} = e^{- R F} \non \\
&=& \langle B_{+} | e^{- L H_{P}} | B_{-} \rangle \non \\
&\approx& \langle B_{+} | 0 \rangle  \langle 0 | B_{-} \rangle 
e^{- L E_{0}} \qquad \mbox{ for  } L \rightarrow \infty \,.
\ee
In the first line, Euclidean time evolves along the circumference of
the cylinder, and $H_{+-}$ is the Hamiltonian for the system with
spatial boundary conditions $\pm$.  In the second line, time evolves
parallel to the axis of the cylinder, $H_{P}$ is the Hamiltonian for
the system with periodic boundary conditions, and $| B_{\pm} \rangle$
are boundary states which encode initial/final (temporal) conditions. 
In the third line, we consider the limit $L \rightarrow \infty$; the
state $| 0 \rangle $ is the ground state of $H_{P}$, and $E_{0}$ is
the corresponding eigenvalue.  The quantity
$\ln \langle B_{+} | 0 \rangle \langle 0 | B_{-} \rangle$ is the
sought-after boundary entropy \cite{AL, LMSS}.  
Taking the logarithm of the above expressions for the partition
function, one obtains
\be
- R F \approx - L E_{0} + 
\ln \langle B_{+} | 0 \rangle  \langle 0 | B_{-} \rangle
\,.
\label{be}
\ee
Whereas the free energy $F$ has a leading contribution which is of
order $L$, we seek here the subleading correction which is of order
$1$.

\subsection{Thermodynamic limit}

We proceed to compute $F$ using the TBA approach
\cite{YY}-\cite{AN}.  To this end, we introduce the densities
$P_{\pm}(\theta)$ of ``magnons'', i.e., of real Bethe Ansatz roots 
$\{ x_{j} \}$ with $\epsilon_{j} = \pm 1$, respectively; and also the 
densities $\rho_{1}(\theta)$ and $\tilde\rho(\theta)$  of particles 
$\{ \theta_{k} \}$ and holes, respectively.  Computing the imaginary
part of the logarithmic derivative of the ``magnonic'' Bethe Ansatz
equations (\ref{BA2}), we obtain \footnote{There is a contribution
$-{1\over 2 \pi L}\Phi(\theta)$ which originates from the 
exclusion \cite{FS, GMN} of the Bethe Ansatz root $x_{j}=0$.}
\be
P_{+}(\theta) + P_{-}(\theta) &=& {1\over 2\pi} \int_{0}^{\infty} 
d\theta'\ \rho_{1}(\theta') \left[ \Phi(\theta-\theta') + 
\Phi(\theta+\theta') \right] \non \\
&+& {1\over 2 \pi L}\left[\Phi(\theta) + \Psi_{\xi_{+}}(\theta) +
\Psi_{\xi_{-}}(\theta) \right] \,,
\ee
where
\be 
\Phi(\theta) &=& {\partial\over \partial \theta} \Im m\ \ln \left(
{\sinh({\theta\over 2} - {i\pi\over 4})\over
 \sinh({\theta\over 2} + {i\pi\over 4})} \right) =
{1\over \cosh \theta} \,, \non  \\
\Psi_{\xi}(\theta) &=& 
{\partial\over \partial \theta} \Im m\ \ln Q(\theta \,, \xi) =
{4 \cos \xi \cosh\theta \over \cos 2 \xi + \cosh 2 \theta} \,.
\label{kernels}
\ee
In the final equality, we have used the expression (\ref{Qx}) for
$Q(\theta \,, \xi)$, and we have assumed that $\xi$ is real.
We present here the results for $N=$ even - Sector II, from which the results 
for the other sectors (see (\ref{calQ})) can be read off by inspection.
Defining $\rho_{1}(\theta)$ for negative values of $\theta$ to be 
equal to $\rho_{1}(|\theta|)$, we obtain the final form
\be
P_{+}(\theta) + P_{-}(\theta) ={1\over 2\pi} \left(
\rho_{1} * \Phi \right)(\theta) + {1\over 2 \pi L}\left[\Phi(\theta)
+ \Psi_{\xi_{+}}(\theta) + \Psi_{\xi_{-}}(\theta) \right] \,,
\label{constraint1}
\ee
where $*$ denotes convolution
\be
\left( f * g \right)(\theta) = \int_{-\infty}^{\infty} 
d\theta'\ f(\theta-\theta')g(\theta') \,.
\ee

Computing the imaginary part of the logarithmic derivative of the Yang
equation (\ref{openYang2}) using the result (\ref{eigenvaluefinal})
for the eigenvalue, we obtain (again for $N=$ even - Sector II)
\be
\rho_{1}(\theta) &+& \tilde\rho(\theta) = {1\over 2 \pi} \Big\{
2 m \cosh \theta + \int_{0}^{\infty}d\theta'\ \rho_{1}(\theta') 
\left[\Phi_{\sigma}(\theta-\theta') + \Phi_{\sigma}(\theta+\theta') \right] 
\non \\
&+& \int_{0}^{\infty}d\theta'\ \Big[ 
P_{+}(\theta')\left( \Phi^{(1)}_{+}(\theta-\theta') 
+ \Phi_{+}^{(2)}(\theta+\theta') \right)  +
P_{-}(\theta')\left( \Phi^{(1)}_{-}(\theta-\theta') 
+ \Phi_{-}^{(2)}(\theta+\theta') \right) \Big] \non \\
&+& {1\over L}\left[-2\gamma -\Phi_{\sigma}(\theta) 
- 2\Phi_{\sigma}(2\theta) 
+ {\partial\over \partial \theta} \Im m\ \ln M(\theta \,, \xi_{+})
+ {\partial\over \partial \theta} \Im m\ \ln M(\theta \,, \xi_{-})
\right] \Big\} \,,
\ee
where
\be
\Phi_{\sigma}(\theta) &=& {\partial\over \partial \theta} \Im m\  \ln \sigma(\theta) 
= {1\over 8}\int_{-\infty}^{\infty} dk\ {e^{i k \theta}\over \cosh^{2} 
{\pi k\over 2}} = {\theta\over 2\pi \sinh \theta} \,, \non  \\
\Phi_{\pm}^{(l)}(\theta) &=& {\partial\over \partial \theta} \Im m\ \ln 
\lambda^{(l)}_{\pm}(\theta)  \,, \quad 
l = 1\,, 2 \,,
\ee
and $\lambda^{(l)}_{\pm}(\theta)$ are introduced in (\ref{lambdas}).
Using the facts $\Phi_{\pm}^{(1)}(\theta) = \Phi_{\pm}^{(2)}(\theta) 
= \pm {1\over 2} \Phi(\theta)$, and
defining $P_{\pm}(\theta)$ for negative values of $\theta$ to be 
equal to $P_{\pm}(|\theta|)$, we obtain
\be
\rho_{1}(\theta) &+& \tilde\rho(\theta) =
{m\over \pi} \cosh \theta 
+ {1\over 2 \pi} \left( \rho_{1} * \Phi_{\sigma}\right) (\theta) 
+ {1\over 4 \pi} \left( (P_{+} - P_{-}) * \Phi \right) (\theta)
\non \\
&+& {1\over 2 \pi L}\left[-2\gamma -\Phi_{\sigma}(\theta) 
- 2\Phi_{\sigma}(2\theta) 
+  {\partial\over \partial \theta}\Im m\ \ln M(\theta \,, \xi_{+})
+  {\partial\over \partial \theta}\Im m\ \ln M(\theta \,, \xi_{-})
\right] \,.
\ee
Using (\ref{constraint1}) to eliminate $P_{-}$, and (\ref{Ptheta}), 
(\ref{Mtheta}) to separate the various factors in $M(\theta \,, \xi)$,
we obtain
\be
\rho_{1}(\theta) + \tilde\rho(\theta) &=& 
{m\over \pi} \cosh \theta 
+ {1\over 2\pi}  P_{+} * \Phi 
+ {1\over 2 \pi} \rho_{1} * \left(\Phi_{\sigma} 
- {1\over 4 \pi} \Phi * \Phi \right) \non \\
&+& {1\over 2 \pi L}\Bigg[ 
2\left({\partial\over \partial \theta} \Im m\ \ln P_{min}(\theta)
- \Phi_{\sigma}(2\theta) - \Phi_{\sigma}(\theta) + \gamma \right) 
+ \left(\Phi_{\sigma} - {1\over 4 \pi} \Phi * \Phi \right) \non \\
&+& \left( \Phi_{\sigma}(\theta - i\xi_{+}) 
+ \Phi_{\sigma}(\theta + i\xi_{+})
- {1\over 4 \pi} \Psi_{\xi_{+}} * \Phi  \right) \non \\
&+& \left( \Phi_{\sigma}(\theta - i\xi_{-}) 
+ \Phi_{\sigma}(\theta + i\xi_{-})
- {1\over 4 \pi} \Psi_{\xi_{-}} * \Phi \right) \non \\
&+& {\partial\over \partial \theta} \Im m\
\ln P^{CDD}(\theta \,, \xi_{+})
+ {\partial\over \partial \theta} \Im m\
\ln P^{CDD}(\theta \,, \xi_{-})
\Bigg] \,. \non \\
\ee
Using the identity \cite{AlZa2}
\be
\Phi_{\sigma}(\theta)  
- {1\over 4 \pi} \left( \Phi * \Phi \right)(\theta)  = 0 \,,
\ee
as well as the identities
\be
\Phi_{\sigma}(\theta - i\xi) + \Phi_{\sigma}(\theta + i\xi)
- {1\over 4 \pi} \left( \Psi_{\xi} * \Phi \right)(\theta) 
&=& 0 \,, \non  \\
{\partial\over \partial \theta}  \Im m\ \ln P_{min}(\theta)
- \Phi_{\sigma}(2\theta) - \Phi_{\sigma}(\theta)
+ \gamma &=& -{1\over 4}\Phi(\theta)  \,, 
\ee 
we arrive at the final simple result
\be
\rho_{1}(\theta) + \tilde\rho(\theta) =
{m\over \pi} \cosh \theta 
+ {1\over 2\pi}  \left( P_{+} * \Phi \right)(\theta)
+ {1\over 2 \pi L}\Bigg[ -{1\over 2}\Phi(\theta) 
+ \kappa_{\xi_{+}}(\theta) + \kappa_{\xi_{-}}(\theta)
\Bigg] \,,
\label{constraint2}
\ee
where 
\be
\kappa_{\xi}(\theta) = {\partial\over \partial \theta} \Im m\
\ln P^{CDD}(\theta \,, \xi) = {4 \sin \xi \cosh \theta \over
\cos 2\xi - \cosh 2\theta } \,.
\label{kappa}
\ee 
The result (\ref{constraint2}) holds in fact for all four sectors.

The thermodynamic limit of the magnonic Bethe Ansatz equations and the
Yang equations, given by (\ref{constraint1}) and (\ref{constraint2}),
respectively, are the main results of this subsection.

\subsection{TBA equations and boundary entropy}

The free energy $F$ is given by
\be
F = E - T S \,,
\ee
where the temperature is $T={1\over R}$, the energy $E$ is
\be
E = \sum_{k=1}^{N} m \cosh \theta_{k} 
= {L\over 2}\int_{-\infty}^{\infty}d\theta\ 
\rho_{1}(\theta) m \cosh \theta \,,
\ee
and the entropy $S$ is \cite{YY, AlZa2}
\be
S &=& {L\over 2}\int_{-\infty}^{\infty}d\theta\ \Big\{
(\rho_{1} + \tilde \rho) \ln (\rho_{1} + \tilde \rho) 
- \rho_{1}  \ln \rho_{1} - \tilde \rho \ln \tilde \rho \non \\
&+& (P_{+} + P_{-}) \ln (P_{+} +  P_{-}) 
- P_{+}  \ln P_{+} -  P_{-} \ln P_{-} \Big\} \,.
\ee
Extremizing the free energy $(\delta F = 0)$ subject to the constraints
\be
\delta P_{-} = -\delta P_{+} 
+ {1\over 2 \pi} \delta \rho_{1} * \Phi \,, \qquad 
\delta \tilde\rho  = -\delta \rho_{1} 
+ {1\over 2 \pi} \delta P_{+} * \Phi \,,
\ee
(which follow from Eqs.  (\ref{constraint1}), (\ref{constraint2}),
respectively) we obtain a set of TBA equations which is the same as
for the case of periodic boundary conditions \cite{AlZa2}
\be
\rr \cosh \theta &=& \epsilon_{1}(\theta) 
+ {1\over 2 \pi} \left( \Phi * L_{2} \right)(\theta) \,, \non  \\
0 &=& \epsilon_{2}(\theta) 
+ {1\over 2 \pi} \left( \Phi * L_{1} \right)(\theta) \,, 
\label{TBA}
\ee
where 
\be
L_{i}(\theta) &=& \ln \left( 1 + e^{-\epsilon_{i}(\theta)} \right) \,,
\qquad \rr = m R \,, \non \\
\epsilon_{1} &=& \ln \left( {\tilde \rho\over \rho_{1}} \right) \,, 
\qquad 
\epsilon_{2} = \ln \left( {P_{-}\over P_{+}} \right) \,.
\ee

We next evaluate $F$ using also the constraints (\ref{constraint1}),
(\ref{constraint2}) and the TBA equations.  From the boundary (order
$1$) contribution, we obtain (see Eq.  (\ref{be})) the boundary
entropy \footnote{Taking into account all the sectors, the last term 
in (\ref{beboth}) should be replaced by $\left[\Phi(\theta)  + 
\Psi^{(\alpha)}(\theta \,, \xi_{\pm}) \right] L_{2}(\theta)$,
where $\alpha$ runs over the four sectors (\ref{sectors}), and 
$\Psi^{(\alpha)}(\theta \,, \xi_{\pm})$ is given by
\be
\Psi^{(\alpha)}(\theta \,, \xi_{\pm}) =
\left\{ \begin{array}{l}
          0 \\
	  \Psi_{\xi_{+}}(\theta) + \Psi_{\xi_{-}}(\theta) \\
	  \Psi_{\xi_{-}}(\theta) \\
	   \Psi_{\xi_{+}}(\theta) \end{array}\right. \,, 
\label{bebothL2}	   
\ee
respectively.}
\be
\ln \langle B_{+} | 0 \rangle  \langle 0 | B_{-} \rangle &=&
{1\over 4\pi}\int_{-\infty}^{\infty}d\theta\ \Big\{ 
\left[ -{1\over 2}\Phi(\theta) 
+ \kappa_{\xi_{+}}(\theta)+ \kappa_{\xi_{-}}(\theta) 
\right] L_{1}(\theta) \non \\
&+& \left[\Phi(\theta)  + 
\Psi_{\xi_{+}}(\theta) + \Psi_{\xi_{-}}(\theta)
\right] L_{2}(\theta) \Big\}
\,. \label{beboth}
\ee
In particular, the dependence of the boundary entropy of a single
boundary on the boundary parameter $\xi$ is given by
\be
\ln g(\xi) =
{1\over 4\pi}\int_{-\infty}^{\infty}d\theta\ 
\left[ \kappa_{\xi}(\theta)\ L_{1}(\theta)  
+ \Psi_{\xi}(\theta)\ L_{2}(\theta) \right] \,,
\label{beresult}
\ee
where the kernels $\kappa_{\xi}(\theta)$ and $\Psi_{\xi}(\theta)$ are
given in Eqs.  (\ref{kappa}) and (\ref{kernels}), respectively.  This
expression for the boundary entropy for the NS case of the TIM is
another of the main results of this paper.

\subsection{Massless boundary flows}

We now consider the case of massless boundary flow.  That is, we
consider the bulk massless scaling limit
\be
m = \mu n\,, \qquad \theta = \hat \theta \mp \ln \frac{n}{2} \,, 
\qquad n \rightarrow 0 \,,
\label{limit}
\ee
where $\mu$ and $\hat \theta$ are finite, which implies $E = \mu
e^{\pm \hat \theta}$, $P = \pm \mu e^{\pm \hat \theta}$.
There are two nontrivial scaling limits of the boundary parameter, 
which we consider in turn. As we shall see, these two limits correspond
to the boundary flows $(-0)\& (0+) \rightarrow (-)\& (+)$ and 
$(-0)\& (0+) \rightarrow 2(0)$, respectively.

\subsubsection{The boundary flow $(-0)\& (0+) \rightarrow (-)\& (+)$}

\begin{figure}[tb]
	\centering
	\epsfxsize=0.80\textwidth\epsfbox{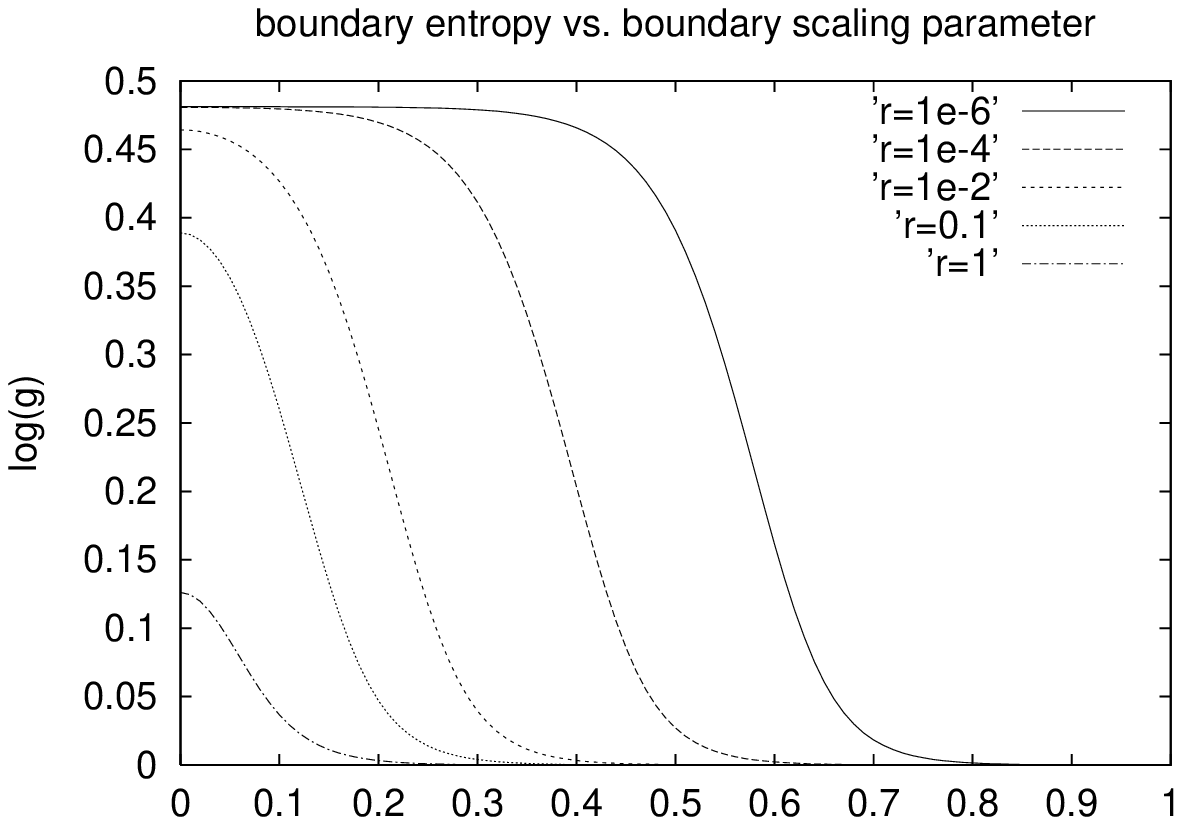}
        \caption[xxx]{\parbox[t]{0.4\textwidth}}
	\label{flow}
\end{figure}

Let us first consider the scaling limit
\be
\xi = - \frac{\pi}{2} + i(\theta_{B} - \ln \frac{n}{2}) \,,
\qquad n \rightarrow 0 \,,
\label{limit1}
\ee
where the boundary scale $\theta_{B}$ is finite. For the sign $-$ in
the limit (\ref{limit}), the CDD factor has a nontrivial limit 
\be
P^{CDD}(\theta \,, \xi) \rightarrow 
-{i\sinh({\hat \theta - \theta_{B}\over 2}- {i \pi\over 4})\over
\sinh({\hat \theta - \theta_{B}\over 2}+ {i \pi\over 4})} \,,
\ee
and therefore, so does the corresponding kernel (\ref{kappa})
\be
\kappa_{\xi}(\theta) \rightarrow \Phi(\hat \theta - \theta_{B}) \,.
\ee
On the other hand, the factor $Q(\theta \,, \xi)$ (\ref{Qx}) becomes 
real in this limit; hence, the corresponding kernel 
$\Psi_{\xi}(\theta)$ (\ref{kernels}) vanishes. The result (\ref{beresult})
for the boundary entropy therefore implies
\be
\ln g = 
{2\over 4\pi}\int_{-\infty}^{\infty}d\hat \theta\ 
\Phi (\hat \theta - \theta_{B})\ \hat L_{1}(\hat \theta)  \,,
\label{flow1}
\ee
where $\hat \epsilon_{i}(\hat \theta) \equiv \epsilon_{i}(\hat \theta -
\ln \frac{n}{2})$, and $\hat L_{i}(\hat\theta) = 
\ln (1 + e^{-\hat \epsilon_{i}(\hat \theta)})$.
The factor of 2 appearing in (\ref{flow1}) accounts for the
contribution from the sign $+$ in the limit (\ref{limit}),
corresponding to the fact that right-movers and left-movers give equal
contributions to the boundary entropy.
In the UV limit $\theta_{B} \rightarrow -\infty$, 
the integrand is nonvanishing for $\hat \theta \rightarrow -\infty$;
similarly, the IR limit $\theta_{B} \rightarrow \infty$ requires
$\hat \theta \rightarrow \infty$. Using the results
$\hat L_{1}(-\infty) =\ln \left( {1\over 2}(3+\sqrt{5})\right)$, 
$\hat L_{1}(\infty)=0$ which follow from the TBA Eqs. (\ref{TBA}), 
we conclude from (\ref{flow1}) that 
\be
{g^{UV}\over g^{IR}} = {1\over 2}(1+\sqrt{5}) \,.
\ee
This is precisely the ratio of $g$ factors corresponding to the
boundary flow $(-0)\& (0+) \rightarrow (-)\& (+)$, as follows
from (\ref{gfactors}),
\be
{g_{(-0)\& (0+)}\over g_{(-)\& (+)}} = 
\eta^{2} = {1\over 2}(1+\sqrt{5}) \label{gratio1}
\,.
\ee
A plot of $\ln g$ in (\ref{beresult}) as a function of the boundary
scaling parameter $\theta_B$ defined in (\ref{limit1}) with finite $n$
\footnote{The horizontal axis is rescaled in such a way that the range
is mapped to $(0 \,, 1)$.}
for various values of $\rr$ is given in Fig.  \ref{flow}.  Observe
that for $\rr << 1$, the correct conformal boundary entropy is
reproduced.  As $\rr$ increases, one can see that the entropy deviates
from the conformal field theory value.  Indeed, for $\rr \rightarrow
1$, the entropy approaches zero, as expected for a massive theory.

\begin{figure}[tb]
	\centering
	\epsfxsize=0.20\textwidth\epsfbox{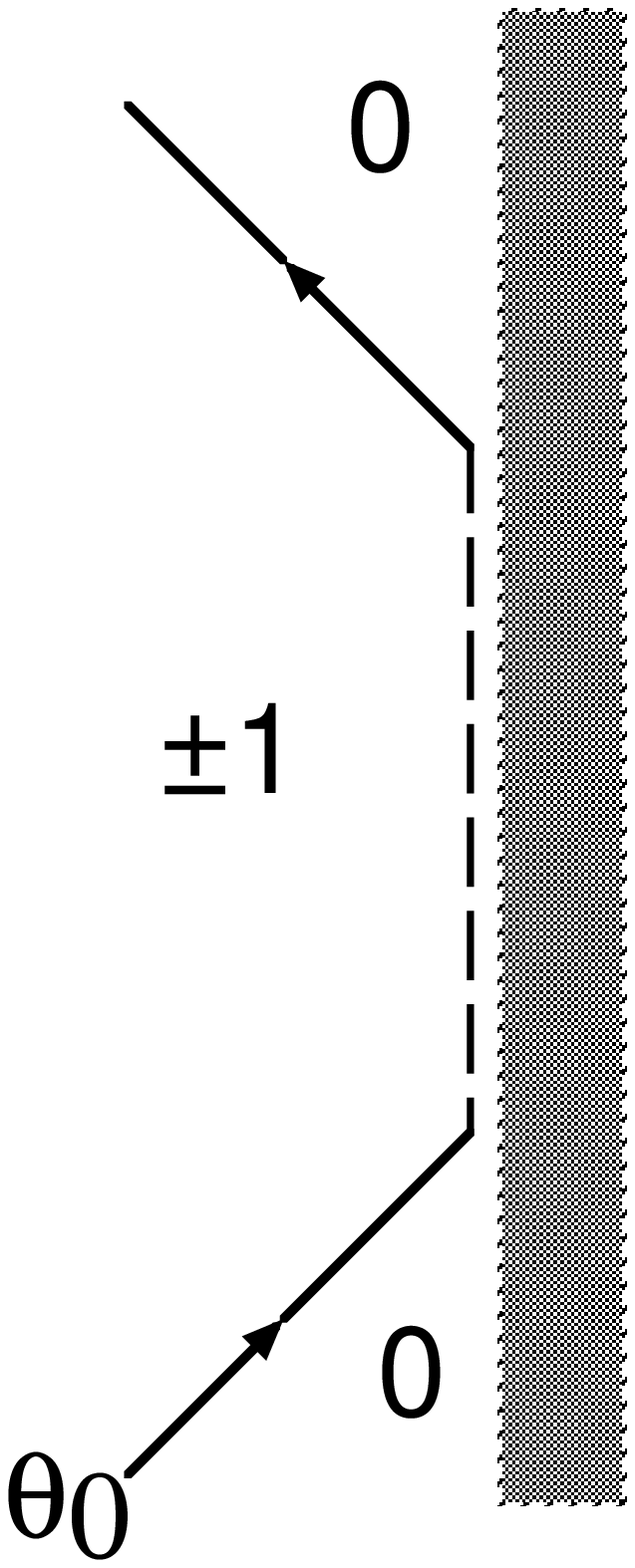}
	\caption[xxx]{\parbox[t]{0.4\textwidth}{
	Boundary bound state pole.}
	}
	\label{fig2a}
\end{figure}

One might wonder how there can be a flow to the boundary condition
$(-)\& (+)$ in the $N=$ even - Sector II, for which the boundary
``spins'' are fixed to 0 (see (\ref{sectors})).  Our explanation is that
there are boundary bound states with spins $\pm 1$, corresponding to 
the pole at $\hat \theta = \theta_{0} \equiv \theta_{B} - {i \pi\over 2}$
in the CDD factor.  (See Figure \ref{fig2a}.)  

\subsubsection{The boundary flow $(-0)\& (0+) \rightarrow 2(0)$}

Let us now consider instead the scaling limit
\be
\xi = i(\theta_{B} - \ln \frac{n}{2}) \,,
\qquad n \rightarrow 0 \,,
\label{limit2}
\ee
with $\theta_{B}$ finite.  Taking again the sign $-$ in the limit
(\ref{limit}), the factor $P^{CDD}(\theta \,, \xi)$ becomes real, and
so the corresponding kernel $\kappa_{\xi}(\theta)$ vanishes.  However,
\be
Q(\theta \,, \xi) \rightarrow 
{i\sinh({\hat \theta - \theta_{B}\over 2}- {i \pi\over 4})\over
\sinh({\hat \theta - \theta_{B}\over 2}+ {i \pi\over 4})} \,,
\ee
and therefore
\be
\Psi_{\xi}(\theta) \rightarrow \Phi(\hat \theta - \theta_{B}) \,.
\ee
The result (\ref{beresult}) for the boundary entropy now implies
\be
\ln g = 
{2\over 4\pi}\int_{-\infty}^{\infty}d\hat \theta\ 
\Phi (\hat \theta - \theta_{B})\ \hat L_{2}(\hat \theta)  \,.
\label{flow2}
\ee
Using the results
$\hat L_{2}(-\infty) =\ln \left( {1\over 2}(3+\sqrt{5})\right)$, 
$\hat L_{2}(\infty)=\ln 2$, 
we obtain \footnote{This flow does not occur for $N=$ even - Sector I,
since for this sector there is no $L_{2}(\theta)$ contribution to the
boundary entropy, as can be seen from (\ref{bebothL2}).  Our
understanding of this fact is as follows: By definition
(\ref{sectors}), this sector has boundary ``spins'' $\pm 1$. 
Moreover, in the scaling limit (\ref{limit2}), there cannot be a
boundary bound state with spin $0$, since the CDD factor does not have
a corresponding pole.  That is, the process represented by Figure
\ref{fig2a} with the spins $\pm 1$ and $0$ interchanged does not
occur.  Hence, there cannot be a flow to the boundary condition $(0)$
in this sector.}
\be
{g^{UV}\over g^{IR}} = {1\over 2}\sqrt{3+\sqrt{5}} \,.
\ee
This is the ratio of $g$ factors corresponding to the
boundary flow $(-0)\& (0+) \rightarrow 2(0)$, since
\be
{g_{(-0)\& (0+)}\over g_{2(0)}} = {\eta^{2}\over \sqrt{2}} 
= {1\over 2}\sqrt{3+\sqrt{5}} \,. \label{gratio2}
\ee

\subsubsection{Nonsupersymmetric flows}\label{subsubsec:nonsusyflows}

The analysis presented so far in Secs.  \ref{sec:inversion} and
\ref{sec:TBA} has been restricted to the NS case of the TIM, for which
the boundary $S$ matrix is given by (\ref{boundNS1}),
(\ref{boundNS2}).  However, the results for the cases without
supersymmetry can now be obtained with no additional effort.

For definiteness, let us now consider the nonsupersymmetric boundary
$S$ matrix (\ref{boundSnosusy1})-(\ref{Mtheta}).  \footnote{For the
other nonsupersymmetric case (\ref{otherboundSnosusy}), the results
are exactly parallel, with the spins $-1$ interchanged with $+1$.}
The corresponding inversion 
identity is again given by (\ref{inversionid1})-(\ref{inversionid3}),
except the sectors are now given by
\be
N=\mbox{ even } - \mbox{ Sector } I &:& 
a_{1}= a'_{1}= a_{N+1}= a'_{N+1} = -1
\,, \non  \\
N=\mbox{ even } - \mbox{ Sector } II &:& 
a_{1} = a'_{1} = a_{N+1} = a'_{N+1} = 0 
\,, \non  \\
N=\mbox{ odd } - \mbox{ Sector } I &:& 
a_{1}= a'_{1} =-1\,, 
\qquad a_{N+1} = a'_{N+1} = 0 \,, \non  \\
N=\mbox{ odd } - \mbox{ Sector } II &:& a_{1} = a'_{1} = 0 \,, 
\qquad a_{N+1} = a'_{N+1} =-1 \,.
\label{nosusysectors}
\ee 
That is, the sectors are restrictions of those in the NS case
(\ref{sectors}).  In particular, the $N=$ even - Sector II is 
identical to the one for the NS case.  Hence, the TBA equations and
boundary entropy are the same as before (\ref{TBA}), (\ref{beresult}). 
Moreover, the two massless scaling limits give the same results
(\ref{flow1}), (\ref{flow2}).  However, the interpretation of these
scaling limits is different from the interpretation in the NS case:
the first scaling limit now corresponds to the boundary flow $(-0)
\rightarrow (-)$, while the second scaling limit now corresponds to
the boundary flow $(-0) \rightarrow (0)$.  That both interpretations
are possible is due to the coincidence in the ratio of $g$ factors
\cite{Ne1},
\be
{g_{(-0)\& (0+)}\over g_{(-)\& (+)}} = {g_{(-0)}\over g_{(-)}} 
\,, \qquad 
{g_{(-0)\& (0+)}\over g_{2(0)}} = {g_{(-0)}\over g_{(0)}}  \,.
\label{gratio3}
\ee
The TBA results (\ref{flow1}), (\ref{flow2}) for these flows coincide
with those obtained in \cite{LSS} on the basis of an analogy with the 
Kondo problem.

\section{R case}\label{sec:Rcase}

We now consider the R case of the TIM, for which
the boundary $S$ matrix is given by (\ref{boundR1})-(\ref{NRtheta}).
Remarkably, the results are closely related (in fact, dual)
to those for the NS case. Indeed, let us define the four sectors
as before (\ref{sectors}). The relation between the full and 
reduced transfer matrices is again given by (\ref{fullreducedreltn1}),
except $w^{(\alpha)}(\theta)$ is now given by
\be
w^{(\alpha)}(\theta) =
\prod_{j=1}^{N}\sigma(\theta-\theta_{j}) \sigma(\theta+\theta_{j}) 
\times
\left\{ \begin{array}{l}
          N(i\pi-\theta \,, \xi_{+})\ N(\theta \,, \xi_{-}) \\
	  R(i\pi-\theta \,, \xi_{+})\ R(\theta \,, \xi_{-})\\
	  e^{-2i \gamma \theta} N(i\pi-\theta \,, \xi_{+})\ 
	  R(\theta \,, \xi_{-}) \\
	  e^{2i \gamma \theta} R(i\pi-\theta \,, \xi_{+})\ 
	  N(\theta \,, \xi_{-}) \end{array}\right. \,.
\ee
The inversion identity is again given by 
(\ref{inversionid1}), (\ref{inversionid2}), except 
$f^{(\alpha)}_{\pm}(\theta)$ is now given by
\be
f^{(\alpha)}_{\pm}(\theta) =
\left\{ \begin{array}{c}
        {1\over 4} (\cosh \theta \pm \cos \xi_{-}) 
	(\cosh \theta \pm \cos \xi_{+})\\
         4\\
	{1\over 2} (\cosh \theta \pm \cos \xi_{+})\\
	 2 (\cosh \theta \pm \cos \xi_{-})
	   \end{array}\right. \,. \label{Rfalpha}
\ee
The periodicity and crossing properties of the reduced transfer
matrix are the same as before (\ref{periodicity}),
(\ref{transfercrossing}).  In contrast to the NS case, the reduced
transfer matrix now becomes an anti-diagonal (rather than diagonal)
matrix for $\theta \rightarrow \infty$.  Nevertheless, the asymptotic
values of the eigenvalues are again given by (\ref{asymptoticeigen}),
except $z^{(\alpha)}$ is now given by
\be
z^{(\alpha)}(\theta) = \left\{ \begin{array}{l}
            \pm  \left(-{i e^{\theta}\over 4}\right)^{{N\over 2}+1}\\
            \pm 2\left(-{i e^{\theta}\over 4}\right)^{N\over 2}\\
	    \pm  \left(-{i e^{\theta}\over 4}\right)^{N+1\over 2}\\
            \pm 2\left(-{i e^{\theta}\over 4}\right)^{N+1\over 2}
	   \end{array}\right. 
\,. 
\ee 
A suitable Ansatz for the eigenvalues is again given by (\ref{Ansatz}),
except $c^{(\alpha)}$ and $d^{(\alpha)}$ are now given by
\be
c^{(\alpha)} = \left\{ \begin{array}{c}
          \pm 1\\
	  \pm 2\\
	  \pm 1\\
          \pm 2\end{array}\right. 
\,, \qquad 
d^{(\alpha)} = \left\{ \begin{array}{c}
          {N\over 2}+ 1\\
	  {N\over 2}\\
          {N+ 1\over 2}\\
          {N+ 1\over 2}\end{array}\right. 
\,,
\ee 
respectively. The Bethe Ansatz equations are therefore again given 
by (\ref{BA1}), with the new $f^{(\alpha)}_{\pm}(\theta)$ given in
(\ref{Rfalpha}). Comparing with the old 
$f^{(\alpha)}_{\pm}(\theta)$ given in (\ref{inversionid3}),
we conclude that the Bethe Ansatz equations for the R case
exactly coincide with those for the NS case, except the sectors I and 
II are interchanged (for both $N=$ even and $N=$ odd)! We remark
that the eigenvalues do not depend on the parameter $r$ which appears 
in the boundary $S$ matrix.

It is now straightforward to repeat the TBA analysis. 
For $N=$ even - Sector I, we obtain the same constraint equations
(\ref{constraint1}), (\ref{constraint2}), and therefore the same
TBA equations (\ref{TBA}) and boundary entropy (\ref{beresult}).
The result (\ref{flow1}) for the first massless scaling limit can now
be interpreted as the boundary flow $(d) \rightarrow (0)$, since
\be
{g_{(d)}\over g_{(0)}} = \eta^{2} = {1\over 2}(1+\sqrt{5}) 
\label{gratio4}
\,,
\ee
as follows from (\ref{gfactors}).  Similarly, the result (\ref{flow2})
for the second massless scaling limit can now be interpreted as the
boundary flow $(d) \rightarrow (-)\& (+)$, since
\be
{g_{(d)}\over g_{(-)\& (+)}} = {\eta^{2}\over \sqrt{2}} 
= {1\over 2}\sqrt{3+\sqrt{5}} \,.
\label{gratio5}
\ee

\section{Conclusion}\label{sec:conclusion}

We have achieved the principal goals set out in the Introduction:
\begin{itemize}
    
\item We have provided support for the proposed TIM boundary $S$
matrices \cite{Ch, Ne1} by showing that the corresponding boundary
entropies (\ref{beresult}), (\ref{flow1}), (\ref{flow2})
are consistent with boundary flows (both supersymmetric
(\ref{gratio1}), (\ref{gratio2}), (\ref{gratio4}), (\ref{gratio5}) and
nonsupersymmetric (\ref{gratio3})) which were expected on other
grounds \cite{Ch, LSS, Af, GRW, FPR, Ne1}.

\item We have developed in Appendix \ref{app:fusion} analytical tools
for RSOS models with boundary, which we have used to derive exact
inversion identities for the TIM. (See
(\ref{inversionid1})-(\ref{inversionid3}) and (\ref{Rfalpha}) for the
supersymmetric cases, and (\ref{nosusysectors}) for the
nonsupersymmetric case.)

\item Our TBA descriptions of boundary flows have been
derived directly from the TIM scattering theory.  The fact that we
have reproduced the TBA description of the nonsupersymmetric flows
given by Lesage {\it et al.} \cite{LSS} provides support for their
approach based on an analogy with the Kondo problem.  The TBA
descriptions of the supersymmetric boundary flows are new.

\end{itemize}

It would be interesting to see if the approach presented here can also
be used to investigate massless flow in the bulk \cite{LSS}. 
Moreover, we expect that it should be possible to generalize this
approach to more complicated models, such as the RSOS${}_{n}$ models
with $n>3$ \cite{AlZa2, LSS}, and coset models \cite{AlZa3, AR}.

\section*{Acknowledgments}

Each of the authors is grateful for the hospitality extended to 
him at the other's home institution.  This work was supported
in part by KOSEF R01-1999-00018 and KRF 2001-015-D00071 (C.A.) 
and by the National Science Foundation under Grants PHY-9870101 
and PHY-0098088 (R.N.).

\appendix

\section{Properties of $S$ matrices}\label{app:props}

We collect here some important properties which are satisfied by the TIM
bulk and boundary $S$ matrices.  

\subsection{Bulk $S$ matrix}

The bulk $S$ matrix (\ref{bulkS1}) - (\ref{sigma}) has
the following symmetries in its indices
\be
S_{a\ b}^{c\ d}(\theta) = S_{b\ a}^{c\ d}(\theta) = S_{a\ b}^{d\ c}(\theta)
\,. \label{symm}
\ee
It also satisfies the crossing relation
\be
S_{a\ b}^{c\ d}(\theta) = S_{c\ d}^{a\ b}(i\pi - \theta) 
\label{crossing}
\ee
and the unitarity relation
\be
\sum_{d} S_{a\ b}^{c\ d}(\theta)\ S_{a\ b}^{d\ c'}(-\theta)
= \delta_{c\ c'} A_{a\ c}\ A_{b\ c} \,,
\label{unitarity}
\ee
where $A_{a\ b}$ is the so-called adjacency matrix (see e.g. 
\cite{BPO})
\be
A_{a\ b} = \delta_{a \,, b-1} + \delta_{a \,, b+1} \,.
\label{adjacency}
\ee
Moreover, it satisfies the Yang-Baxter (star-triangle) equation
\be
\lefteqn{\sum_{g} 
S_{a\ c}^{b\ g}(\theta_{1} - \theta_{2})\
S_{g\ d}^{c\ e}(\theta_{1} - \theta_{3})\ 
S_{a\ e}^{g\ f}(\theta_{2} - \theta_{3})} \non  \\
&=& \sum_{g} 
S_{b\ d}^{c\ g}(\theta_{2} - \theta_{3})\ 
S_{a\ g}^{b\ f}(\theta_{1} - \theta_{3})\
S_{f\ d}^{g\ e}(\theta_{1} - \theta_{2})
\,. \label{YBE}
\ee
Finally, we note that the bulk $S$ matrix at zero rapidity is
given by
\be
S_{a\ b}^{c\ d}(0) = \delta_{c \,, d} A_{a\ c}\ A_{b\ d}
\,. \label{zerorapid}
\ee

\subsection{Boundary $S$ matrix}

The non-supersymmetric boundary $S$ matrix 
(\ref{boundSnosusy1})-(\ref{Mtheta}) obeys the 
unitarity relation
\be
\sum_{c}
\R{c}{a}{b}(\theta)\ \R{d}{a}{c}(-\theta)
= \delta_{b \,, d}\ A_{a\ b}\ B_{d} \,,
\label{boundunitarity}
\ee 
where here $B_{d}$ equals $1$ if $d$ is an allowed state of the 
boundary and equals zero otherwise; hence, it is given by 
\be
B_{d} = \delta_{d \,, -1} + \delta_{d\,, 0}\,.
\ee
This $S$ matrix also obeys the boundary crossing-unitarity relation
\cite{GZ}
\be
\R{a}{b}{c}({i \pi\over 2}-\theta) 
= \sum_{d}
S_{a\ c}^{b\ d}(2\theta)\  
\R{a}{d}{c}({i \pi\over 2}+\theta) 
\,, \label{boundcrossunit}
\ee
as well as the boundary Yang-Baxter 
equation \cite{Ch, AK, BPO, Zh, Cherednik}
\be
\lefteqn{\sum_{f \,, g}
S_{a\ c}^{b\ g}(\theta_{1} - \theta_{2})\
\R{f}{g}{c}(\theta_{1})\
S_{a\ f}^{g\ d}(\theta_{1} + \theta_{2})\
\R{e}{d}{f}(\theta_{2})} \non  \\
&=& \sum_{f \,, g}
\R{g}{b}{c}(\theta_{2})\
S_{a\ g}^{b\ f}(\theta_{1} + \theta_{2})\
\R{e}{f}{g}(\theta_{1})\
S_{a\ e}^{f\ d}(\theta_{1} - \theta_{2})
\,. \label{BYBE}
\ee
The supersymmetric boundary $S$ matrices described in Sec. 
\ref{subsubsec:susy} obey the unitarity condition (\ref{boundunitarity}) 
with $B_{d}=1$, and also (\ref{boundcrossunit}), (\ref{BYBE}).

\section{Fusion procedure for RSOS models with 
boundary}\label{app:fusion}

The fusion procedure was developed for bulk vertex models in
\cite{vertex, KS}, and was adapted to bulk RSOS models in \cite{RSOS}. 
The fusion procedure was extended to vertex models with boundary in
\cite{MN}, but this work has been adapted only in part to the RSOS
case \cite{BPO, Zh}.  In particular, the useful notions of projectors
and quantum determinants have not been explicitly implemented in
\cite{BPO, Zh}.  For this reason, and also to make this paper
self-contained, we give here a brief summary of the fusion procedure
for RSOS models with boundary, and provide the derivation of the TIM
inversion identity.  However, our treatment is not completely general. 
In particular, to avoid complications which are not necessary for the
TIM, we restrict to $S$ matrices with the symmetries (\ref{symm}).

We remind the reader that a bar over a quantity (e.g., $\bar S$)
denotes that it is ``reduced,'' and a tilde over a quantity (e.g.,
$\tilde S$) denotes that it is ``fused.'' 

\subsection{Projectors}

We shall carry out the fusion procedure by exploiting the fact that 
the reduced \footnote{Note that we work here with the reduced matrix
$\bar S_{a\ b}^{c\ d}(\theta)$ rather than the full matrix
$S_{a\ b}^{c\ d}(\theta)$.  There are good reasons for so doing: (i) as
explained in Sec.  \ref{sec:inversion}, it is the reduced transfer matrix
for which we require an inversion identity; and (ii) the full $S$ matrix
is singular at $\theta= -i\pi$.}
bulk $S$ matrix degenerates into the projector $P_{\ \ \ a\ b}^{-\ c\
d}$ for some value of the rapidity, which for the TIM is $\theta= -i
\pi$, 
\be
\bar S_{a\ b}^{c\ d}(-i \pi) = \sqrt{2}\ P_{\ \ \ a\ b}^{-\ c\ d}
\,. \label{Sdegeneration}
\ee
For the TIM, $P_{\ \ \ a\ b}^{-\ c\ d}$ has the matrix elements
\be
P_{\ \ \ 0\ 0}^{-\ \sigma\ \sigma'} &=& 
{1\over 2}(\delta_{\sigma\,, \sigma'} - \delta_{\sigma\,, -\sigma'})
\,, \non \\
P_{\ \ \ \sigma\ \sigma'}^{-\ 0\ 0} &=& \delta_{\sigma\,, -\sigma'} 
\,, \label{projpminus}
\ee
where as usual $\sigma \,, \sigma' \in \{ -1 \,, +1 \}$.
We define the projector $P_{\ \ \ a\ b}^{+\ c\ d}$ by
\be
P_{\ \ \ a\ b}^{+\ c\ d} = \id_{a\ b}^{c\ d} - P_{\ \ \ a\ b}^{-\ c\ d}
\,, 
\ee
where $\id_{a\ b}^{c\ d}$ is the ``adjacency-inclusive'' identity matrix,
\be
\id_{a\ b}^{c\ d} = \delta_{c \,, d} A_{a\ c}\ A_{b\ d} \,.
\label{identitymatrix}
\ee 
For the TIM, $P_{\ \ \ a\ b}^{+\ c\ d}$ has the matrix elements
\be
P_{\ \ \ 0\ 0}^{+\ \sigma\ \sigma'} &=& 
{1\over 2}(\delta_{\sigma\,, \sigma'} + \delta_{\sigma\,, -\sigma'}) 
= {1\over 2} \,, \non \\
P_{\ \ \ \sigma\ \sigma'}^{+\ 0\ 0} &=&  \delta_{\sigma\,, \sigma'} 
\,. \label{projpplus}
\ee
These projectors have the important properties
\be
\sum_{c'} P_{\ \ \ a\ b}^{-\ c\ c'}\ P_{\ \ \ a\ b}^{-\ c'\ d} &=&
P_{\ \ \ a\ b}^{-\ c\ d} \,, \qquad 
\sum_{c'} P_{\ \ \ a\ b}^{+\ c\ c'}\ P_{\ \ \ a\ b}^{+\ c'\ d} =
P_{\ \ \ a\ b}^{+\ c\ d} \,, \non \\
\sum_{c'} P_{\ \ \ a\ b}^{-\ c\ c'}\ P_{\ \ \ a\ b}^{+\ c'\ d} &=&  0 \,.
\label{projprops}
\ee

\subsection{Fused bulk $S$ matrices}

\begin{figure}[tb]
	\centering
	\epsfxsize=0.25\textwidth\epsfbox{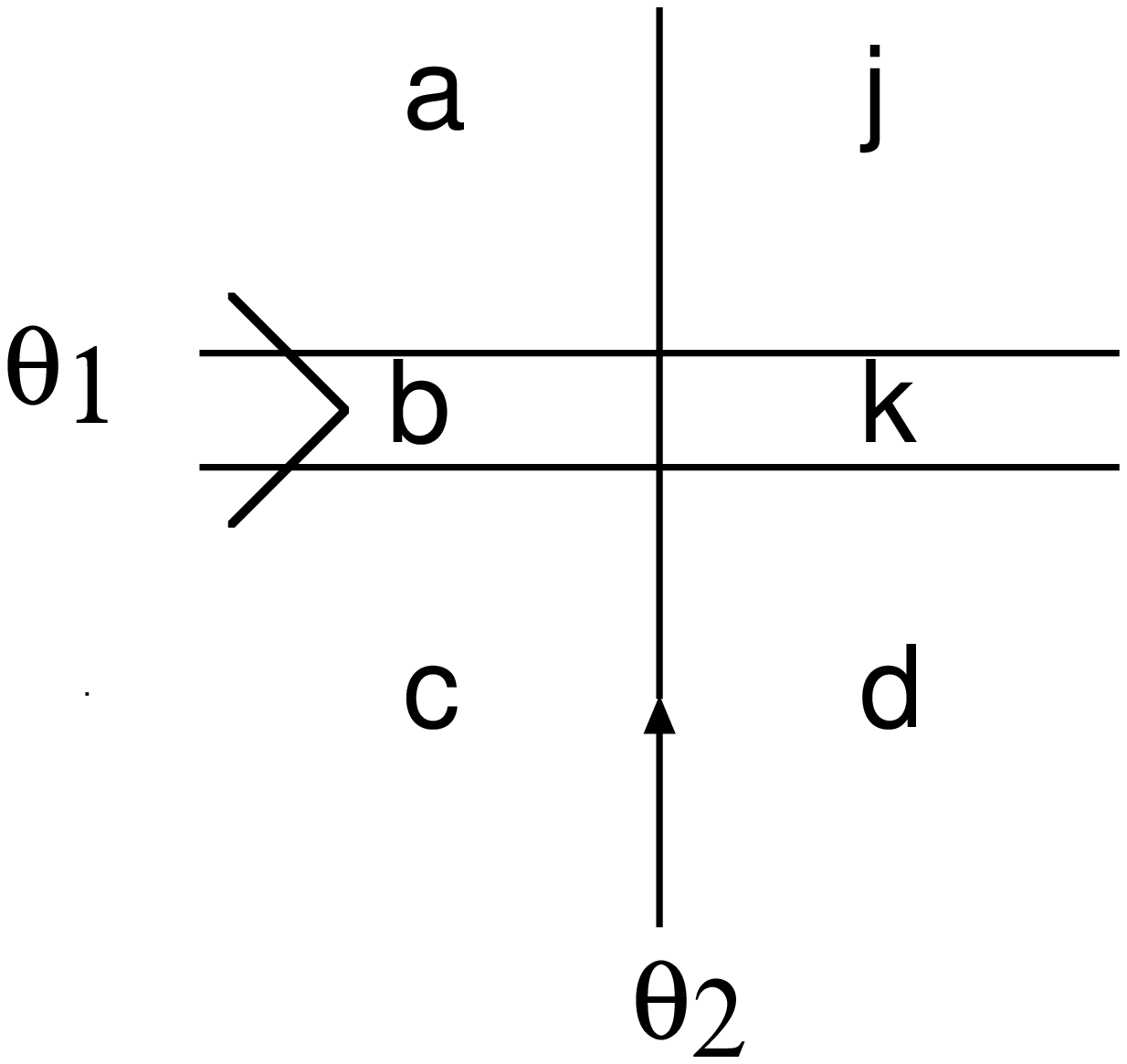}
	\caption[xxx]{\parbox[t]{0.4\textwidth}{
	Fused bulk $S$ matrix $\fusedS{c\ j}{b\ k}{a\ d}
	(\theta_{1}-\theta_{2})$.}
	}
	\label{fig3}
\end{figure}

We derive a bulk ``fusion identity'' from a degeneration of
the Yang-Baxter equation.  That is, in (\ref{YBE}) we set 
$\theta_{1}=\theta$, $\theta_{2}=\theta+ i\pi$, $\theta_{3}=0$, use
the degeneration result (\ref{Sdegeneration}), and contract on the
right of both sides with the projector $P^{+}$ to obtain
\be
\sum_{f \,, g}
P_{\ \ \ a\ c}^{-\ b\ g}\
\bar S_{g\ d}^{c\ f}(\theta)\ 
\bar S_{a\ f}^{g\ j}(\theta+ i\pi)\ 
P_{\ \ \ j\ d}^{+\ f\ k} = 0 \,.
\label{bulkfusionid}
\ee
This identity can be used to show that the ``fused'' $S$ matrix (which 
can be read off from (\ref{bulkfusionid}) by replacing the projector
$P^{-}$ with $P^{+}$, and which is represented by Figure \ref{fig3}), 
\be
\fusedS{c\ j}{b\ k}{a\ d}(\theta) = 
\sum_{f \,, g}
P_{\ \ \ a\ c}^{+\ b\ g}\
\bar S_{g\ d}^{c\ f}(\theta)\ 
\bar S_{a\ f}^{g\ j}(\theta+ i\pi)\ 
P_{\ \ \ j\ d}^{+\ f\ k}
\label{bulkSfused}
\ee
satisfies the generalized Yang-Baxter equation
\be
\lefteqn{\sum_{f\,, g}
\fusedS{b\ f}{k\ g}{a\ c}(\theta_{1} - \theta_{2})\
\fusedS{c\ m}{g\ l}{f\ d}(\theta_{1} - \theta_{3})\
\bar S_{a\ m}^{f\ j}(\theta_{2} - \theta_{3})}\non \\
&=&
\sum_{f\,, g}
\bar S_{b\ d}^{c\ f}(\theta_{2} - \theta_{3})\
\fusedS{b\ j}{k\ g}{a\ f}(\theta_{1} - \theta_{3})\
\fusedS{f\ m}{g\ l}{j\ d}(\theta_{1} - \theta_{2})
\,.
\ee 
For the TIM, the nonzero matrix elements of $\tilde{\bar S}$
are given by
\be
\fusedS{\sigma\ 0}{0\ \sigma'}{\sigma\ 0}(\theta) =
\fusedS{0\ \sigma}{\sigma'\ 0}{0\ \sigma}(\theta) =
{1\over \sqrt{2}}\cosh{\theta\over 2} \,.
\ee

\begin{figure}[tb]
	\centering
	\epsfxsize=0.25\textwidth\epsfbox{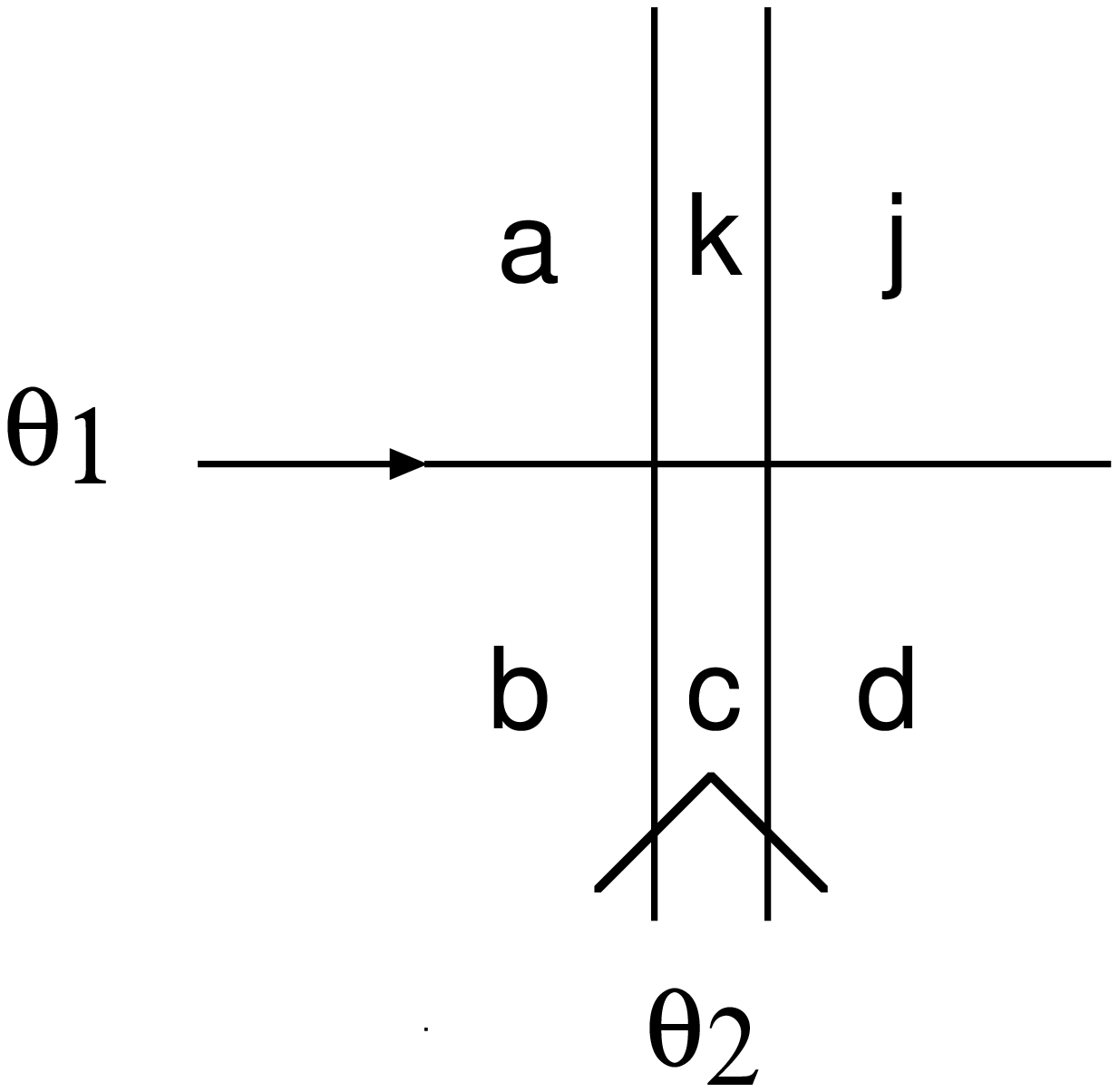}
	\caption[xxx]{\parbox[t]{0.4\textwidth}{
	Fused bulk $S$ matrix $\tilde{\bar S}'^{b\ k\ d}_{a\ c\ j}
	(\theta_{1}-\theta_{2})$.}
	}
	\label{fig4}
\end{figure}

From a second degeneration of the Yang-Baxter equation (\ref{YBE})
with $\theta_{2}-\theta_{3}=-i\pi$, we obtain a second fused
$S$ matrix (see Figure \ref{fig4})
\be
\tilde{\bar S}'^{b\ k\ d}_{a\ c\ j}(\theta)=
\sum_{f\,, g}
P_{\ \ \ b\ d}^{+\ c\ g}\
\bar S_{a\ g}^{b\ f}(\theta-i\pi)\ 
\bar S_{f\ d}^{g\ j}(\theta)\ 
P_{\ \ \ a\ j}^{+\ f\ k} \,,
\label{bulkSfusedprime}
\ee
which obeys 
\be
\lefteqn{\sum_{f\,, g}
\tilde{\bar S}'^{b\ g\ c}_{a\ k\ f}(\theta_{1} - \theta_{2})\
\bar S_{f\ d}^{c\ m}(\theta_{1} - \theta_{3})\
\fusedS{f\ j}{g\ l}{a\ m}(\theta_{2} - \theta_{3})}\non \\
&=&
\sum_{f\,, g}
\fusedS{c\ f}{k\ g}{b\ d}(\theta_{2} - \theta_{3})\
\bar S_{a\ f}^{b\ j}(\theta_{1} - \theta_{3})\
\tilde{\bar S}'^{f\ l\ d}_{j\ g\ m}(\theta_{1} - \theta_{2})
\,.
\ee
For the TIM, the nonzero matrix elements of $\tilde{\bar S}'$
are given by
\be
\tilde{\bar S}'^{0\ 0\ 0}_{\sigma\ \sigma'\ \sigma}(\theta) =
\tilde{\bar S}'^{\sigma\ \sigma'\ \sigma}_{0\ 0\ 0}(\theta) =
{1\over \sqrt{2}}\cosh{\theta\over 2} \,.
\ee

\subsection{Fused boundary $S$ matrix}

\begin{figure}[tb]
	\centering
	\epsfxsize=0.20\textwidth\epsfbox{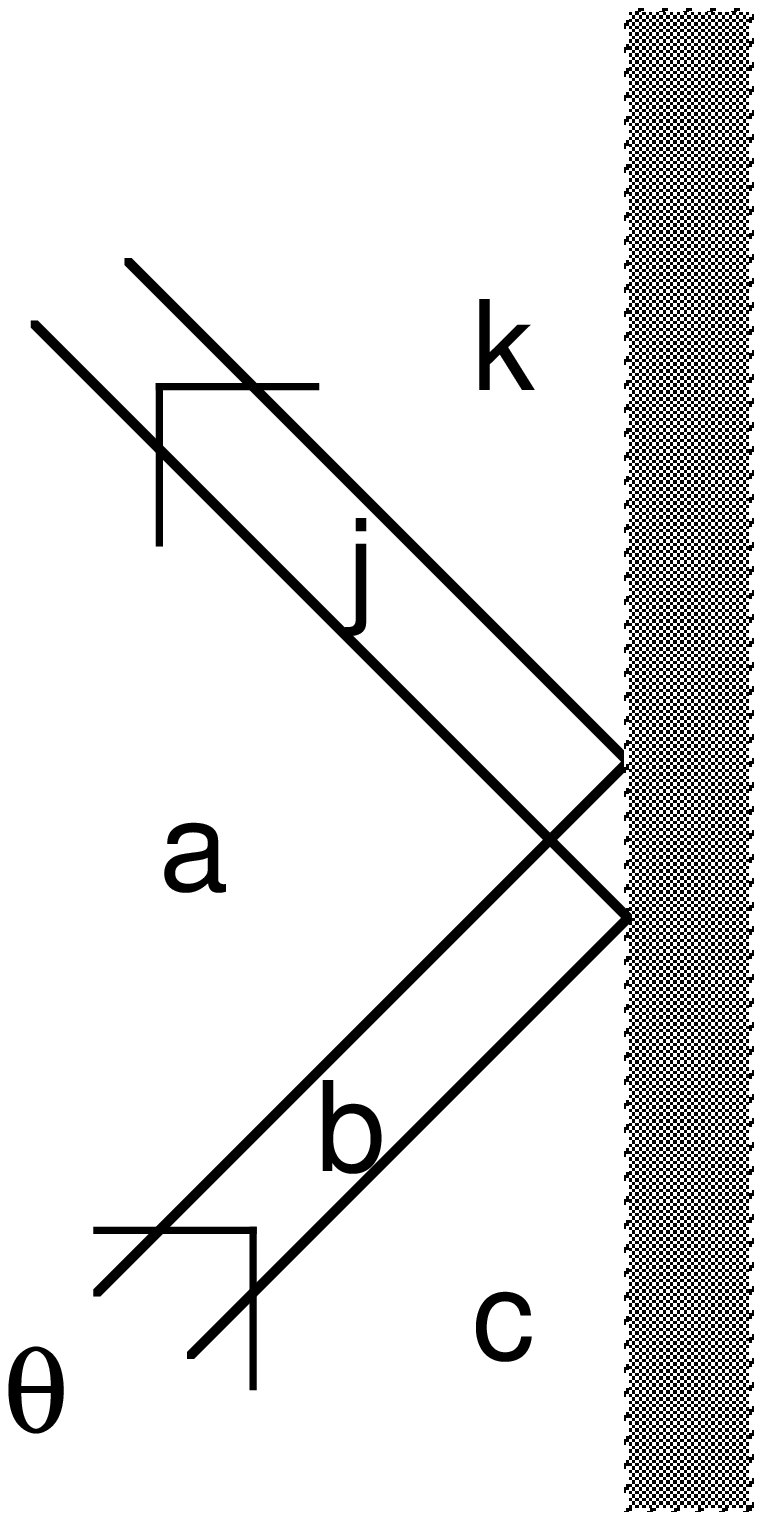}
	\caption[xxx]{\parbox[t]{0.4\textwidth}{
	Fused boundary $S$ matrix $\fusedR{\ \ j\ k}{a\ \ \ }{\ \ b\ c}
        (\theta)$.}
	}
	\label{fig5}
\end{figure}

Following \cite{MN}, we obtain a boundary fusion identity from the
degeneration of the boundary Yang-Baxter equation (\ref{BYBE}) with
$\theta_{1}-\theta_{2}=-i\pi$,
\be
\sum_{d\,, f \,, g}
P_{\ \ \ a\ c}^{-\ b\ g}\
\barR{f}{g}{c}(\theta)\
\bar S_{a\ f}^{g\ d}(2\theta+i\pi)\
\barR{k}{d}{f}(\theta+i\pi)\
P_{\ \ \ a\ k}^{+\ d\ j} = 0 \,.
\label{boundfusionid}
\ee
This identity can be used to show that the ``fused'' $R$ matrix (which 
can be read off from (\ref{boundfusionid}) by replacing the projector
$P^{-}$ with $P^{+}$, and which is represented by Figure \ref{fig5}), 
\be
\fusedR{\ \ j\ k}{a\ \ \ }{\ \ b\ c}(\theta) = \sum_{d\,, f \,, g}
P_{\ \ \ a\ c}^{+\ b\ g}\
\barR{f}{g}{c}(\theta)\
\bar S_{a\ f}^{g\ d}(2\theta+i\pi)\
\barR{k}{d}{f}(\theta+i\pi)\
P_{\ \ \ a\ k}^{+\ d\ j} 
\label{fusedR}
\ee
satisfies the generalized boundary Yang-Baxter equation
\be
\lefteqn{\sum_{f\,, g\,, h\,, i}
\tilde{\bar S}'^{a\ h\ c}_{d\ b\ g}(\theta_{1}-\theta_{2})\
\barR{f}{g}{c}(\theta_{1})\
\fusedS{g\ e}{h\ i}{d\ f}(\theta_{1}+\theta_{2})\
\fusedR{\ \ j\ k}{e\ \ \ }{\ \ i\ f}(\theta_{2})}\non \\
&=&
\sum_{f\,, g\,, h\,, i}
\fusedR{\ \ i\ g}{a\ \ \ }{\ \ b\ c}(\theta_{2})
\tilde{\bar S}'^{a\ h\ g}_{d\ i\ f}(\theta_{1}+\theta_{2}+i\pi)\
\barR{k}{f}{g}(\theta_{1})\
\fusedS{f\ e}{h\ j}{d\ k}(\theta_{1}-\theta_{2}-i\pi) \,.
\ee
The shifts in the arguments of the fused bulk $S$ matrices on the
RHS should be noted.
As an example, for the $NS$ case of the TIM, the nonzero matrix
elements of $\tilde {\bar R}$ are given by
\be
\fusedR{\ \ 0\ \sigma}{\sigma\ \ \ }{\ \ 0\ \sigma}(\theta \,, \xi) = 
\sqrt{2} \cosh{\theta\over 2} \,, \qquad 
\fusedR{\ \ \sigma\ 0}{0\ \ \ }{\ \ \sigma'\ 0}(\theta\,, \xi) = 
{1\over 2\sqrt{2}}(\cosh \theta + \cos \xi) \cosh{\theta\over 2}
\,.
\ee 

\subsection{Fused transfer matrix}

Before attempting to construct the fused transfer matrix, it is 
instructive to first review the construction of the fundamental
transfer matrix. To this end, we set
\be
\bar t_{a'_{1} \ldots a'_{N+1}}^{a_{1} \ldots a_{N+1}}
(\theta | \theta_{1} \,, \ldots \,, \theta_{N})
= \sum_{a''_{1}} 
\barRplus{a_{1}}{a''_{1}}{a'_{1}}(\theta- i \pi\,, \xi_{+})\ 
\scriptT{a_{1}\ldots a_{N+1}}{a''_{1}}{a'_{1}\ldots a'_{N+1}}
(\theta | \theta_{1} \,, \ldots \,, \theta_{N}) \,,
\label{fundamtransfer}
\ee
where $\bar {\cal T}$ is defined by
\be 
\scriptT{a_{1}\ldots a_{N+1}}{a''_{1}}{a'_{1}\ldots a'_{N+1}}
(\theta | \theta_{1} \,, \ldots \,, \theta_{N}) &=&
\sum_{a''_{2}\,, \ldots \,, a''_{N+1}} \Big\{
\monodromy{a''_{1}\ldots a''_{N+1}}{a'_{1}\ldots a'_{N+1}}
(\theta | \theta_{1} \,, \ldots \,, \theta_{N})\
\barR{\ a_{N+1}}{a''_{N+1}}{\ a'_{N+1}}(\theta \,, \xi_{-}) \non \\
&\times&
\hatmonodromy{a_{1}\ldots a_{N+1}}{a''_{1}\ldots a''_{N+1}}
(\theta | \theta_{1} \,, \ldots \,, \theta_{N}) \Big\}\,,
\label{calT}
\ee
and the monodromy matrices $\bar T$ and $\hat{\bar T}$ are given by
\be
\monodromy{a_{1}\ldots a_{N+1}}{a'_{1}\ldots a'_{N+1}}
(\theta | \theta_{1} \,, \ldots \,, \theta_{N}) &=& 
S_{a_{1}\ a'_{2}}^{a'_{1}\ a_{2}}(\theta-\theta_{1}) \ldots
S_{a_{N}\ a'_{N+1}}^{a'_{N}\ a_{N+1}}(\theta-\theta_{N}) \,, \non  \\ 
\hatmonodromy{a_{1}\ldots a_{N+1}}{a'_{1}\ldots a'_{N+1}}
(\theta | \theta_{1} \,, \ldots \,, \theta_{N}) &=& 
S_{a'_{N}\ a_{N+1}}^{a'_{N+1}\ a_{N}}(\theta +\theta_{N}) \ldots
S_{a'_{1}\ a_{2}}^{a'_{2}\ a_{1}}(\theta+\theta_{1}) \,.
\ee
The boundary matrix $\bar R$ in (\ref{calT}) is assumed to obey the 
boundary Yang-Baxter equation (\ref{BYBE}), which implies that
$\bar {\cal T}$ obeys 
\be
\lefteqn{\sum_{c_{1}\,, b_{1}\,, \ldots \,, b_{N+1}}
\bar S^{e\ c_{1}}_{a''_{1}\ a'_{1}}(\theta_{1}-\theta_{2})\
\scriptT{b_{1}\ldots b_{N+1}}{c_{1}}{a'_{1}\ldots a'_{N+1}}(\theta_{1})\
\bar S^{c_{1}\ d_{1}}_{a''_{1}\ b_{1}}(\theta_{1}+\theta_{2})\
\scriptT{a_{1}\ldots a_{N+1}}{d_{1}}{b_{1}\ldots b_{N+1}}(\theta_{2})}
\non \\
&=& \sum_{c_{1}\,, b_{1}\,, \ldots \,, b_{N+1}}
\scriptT{b_{1}\ldots b_{N+1}}{e}{a'_{1}\ldots a'_{N+1}}(\theta_{2})\
\bar S^{e\ c_{1}}_{a''_{1}\ b_{1}}(\theta_{1}+\theta_{2})\
\scriptT{a_{1}\ldots a_{N+1}}{c_{1}}{b_{1}\ldots b_{N+1}}(\theta_{1})\
\bar S^{c_{1}\ d_{1}}_{a''_{1}\ a_{1}}(\theta_{1}-\theta_{2}) \,.
\label{calTYBE}
\ee 
However, the matrix $\bar R^{+}$ in (\ref{fundamtransfer}) is not yet
specified.  Indeed, following Sklyanin \cite{Sk}, the requirement that
the transfer matrix obey the commutativity property
(\ref{opencommutativity}) determines the relation which $\bar R^{+}$
should satisfy.  In this way, we find (using also the properties 
(\ref{symm})-(\ref{unitarity})) that
\be
\barRplus{c}{a}{b}(\theta \,, \xi_{+}) 
= \barR{c}{a}{b}(-\theta \,, \xi_{+}) \,,
\label{boundRplus}
\ee
where $\barR{c}{a}{b}(\theta \,, \xi)$  obeys (\ref{BYBE}).
The result (\ref{fundamtransfer}), (\ref{boundRplus}) coincides with
the expression (\ref{transfer}) for the (reduced) fundamental
open-chain transfer matrix .

We follow a similar strategy to construct the fused transfer matrix
$\tilde{\bar t}$. We set
\be
\tilde{\bar t}_{a'_{1} \ldots a'_{N+1}}^{a_{1} \ldots a_{N+1}}
(\theta | \theta_{1} \,, \ldots \,, \theta_{N})
= \sum_{a''_{1}\,, b''_{1}\,, c''_{1}} 
\fusedRplus{a_{1}\ c''_{1}\ \ }{\quad a''_{1}}{a'_{1}\ b''_{1}\ \ }
(\theta- i \pi\,, \xi_{+})\ 
\scriptTfused{a_{1}\ldots a_{N+1}}{a''_{1}\,, b''_{1}\,, c''_{1}}
{a'_{1}\ldots a'_{N+1}}
(\theta | \theta_{1} \,, \ldots \,, \theta_{N}) \,,
\label{fusedtransfer}
\ee
where $\tilde{\bar{\cal T}}$ is defined by
\be
\lefteqn{\scriptTfused{a_{1}\ldots a_{N+1}}{a''_{1}\,, b''_{1}\,, c''_{1}}
{a'_{1}\ldots a'_{N+1}}
(\theta | \theta_{1} \,, \ldots \,, \theta_{N}) =
\sum_{a''_{2}\,, \ldots \,, a''_{N+1} \,, \atop b''_{N+1} \,, c''_{N+1}} 
\Big\{
\fusedmonodromy{a''_{1}\ldots a''_{N+1}}{b''_{1} \,, b''_{N+1}}
{a'_{1}\ldots a'_{N+1}}
(\theta | \theta_{1} \,, \ldots \,, \theta_{N})} \non \\
&\times&\fusedR{\ \ \ c''_{N+1}\ a_{N+1}}{a''_{N+1}\ \ \ }
{\ \  \ b''_{N+1}\ a'_{N+1}}(\theta \,, \xi_{-})\
\fusedhatmonodromy{a_{1}\ldots a_{N+1}}{c''_{1} \,, c''_{N+1}}
{a''_{1}\ldots a''_{N+1}}
(\theta + i\pi| \theta_{1} \,, \ldots \,, \theta_{N}) \Big\}\,,
\label{fusedcalT}
\ee
where the fused monodromy matrices $\tilde {\bar T}$ and 
$\tilde{\hat{\bar T}}$ are given by
\be
\fusedmonodromy{a_{1}\ldots a_{N+1}}{b_{1} \,, b_{N+1}}
{a'_{1}\ldots a'_{N+1}}
(\theta | \theta_{1} \,, \ldots \,, \theta_{N}) &=&
\sum_{b_{2}\,, \ldots \,, b_{N}}
\fusedS{a'_{1}\ a_{2}}{b_{1}\ b_{2}}{a_{1}\ a'_{2}}(\theta-\theta_{1})
\ldots 
\fusedS{a'_{N}\ a_{N+1}}{b_{N}\ b_{N+1}}{a_{N}\ a'_{N+1}}(\theta-\theta_{N})
\,, \non \\ 
\fusedhatmonodromy{a_{1}\ldots a_{N+1}}{b_{1} \,, b_{N+1}}
{a'_{1}\ldots a'_{N+1}}
(\theta| \theta_{1} \,, \ldots \,, \theta_{N}) &=&
\sum_{b_{2}\,, \ldots \,, b_{N}}
\tilde {\bar S}'^{a'_{N+1}\ b_{N}\ a_{N+1}}_{a'_{N}\ b_{N+1}\ a_{N}} 
(\theta+\theta_{N}) \ldots
\tilde {\bar S}'^{a'_{2}\ b_{1}\ a_{2}}_{a'_{1}\ b_{2}\ a_{1}} 
(\theta+\theta_{1})\,.
\label{fusedmondoromy}
\ee 

We determine the relation obeyed by $\tilde {\bar R}^{+}$ from the
requirement that the fused transfer matrix (\ref{fusedtransfer})
commute with the fundamental transfer matrix (\ref{fundamtransfer}),
(\ref{boundRplus}),
\be
\left[ t(\theta | \theta_{1} \,, \ldots \,, \theta_{N}) \,,
\tilde{\bar t}(\theta' | \theta_{1} \,, \ldots \,, \theta_{N})  
\right] = 0 \,.
\label{fusedcommutativity}
\ee
With the help of the relation obeyed by $\tilde{\bar{\cal T}}$
\be
\lefteqn{\sum_{g\,, h\,, i\,, \atop a'_{1}\,, \ldots \,, a'_{N+1}}
\tilde {\bar S}'^{a\ h\ a_{1}}_{d\ b\ g}(\theta_{1}-\theta_{2})\
\scriptT{a'_{1}  \ldots  a'_{N+1}}{g}{a_{1} \ldots a_{N+1}}(\theta_{1})\
\fusedS{g\  e}{h\ i}{d\ a'_{1}}(\theta_{1}+\theta_{2})\
\scriptTfused{b_{1} \ldots b_{N+1}}{e \,, i \,, j}{a'_{1} \ldots a'_{N+1}}
(\theta_{2})} \non \\
&=&
\sum_{g\,, h\,, i\,, \atop a'_{1}\,, \ldots \,, a'_{N+1}}
\scriptTfused{a'_{1} \ldots a'_{N+1}}{a \,, b \,, i}{a_{1} \ldots a_{N+1}}
(\theta_{2})\
\tilde {\bar S}'^{a\ h\ a'_{1}}_{d\ i\ f}(\theta_{1}+\theta_{2}+i\pi)\
\scriptT{b_{1} \ldots b_{N+1}}{f}{a'_{1} \ldots a'_{N+1}}(\theta_{1})\
\fusedS{f\  e}{h\ j}{d\ b_{1}}(\theta_{1}-\theta_{2}-i\pi) \,, \non \\
\ee 
we obtain the following equation for $\tilde {\bar R}^{+}$
\be
\lefteqn{\sum_{b\,, d\,, f \,, g}
\fusedS{g\ k}{f\ l}{j\ m}(\theta_{2}-\theta_{1})\
\barRplus{d}{g}{m}(\theta_{1}-i\pi \,, \xi_{+})\
\tilde {\bar S}'^{a\ f\ d}_{j\ b\ g}(-\theta_{1}-\theta_{2}+2i\pi)\
\fusedRplus{e\ c\ \ }{\quad a}{d\ b\ \ }
(\theta_{2}- i \pi\,, \xi_{+})} \non \\
&=&
\sum_{b\,, d\,, f \,, g}
\fusedRplus{d\ b\ \ }{\quad k}{m\ l\ \ }
(\theta_{2}- i \pi\,, \xi_{+})\ 
\fusedS{g\ k}{f\ b}{j\ d}(-\theta_{1}-\theta_{2}+i\pi)\
\barRplus{e}{g}{d}(\theta_{1}-i\pi \,, \xi_{+})\
\tilde {\bar S}'^{a\ f\ e}_{j\ c\ g}(\theta_{2}-\theta_{1}+i\pi)\
\,.\non \\
\ee
That is, this relation guarantees the commutativity 
(\ref{fusedcommutativity}). This relation is satisfied by
\be
\fusedRplus{k\ j\ \ }{\quad a}{c\ b\ \ } (\theta- i \pi\,, \xi_{+})
= \sum_{d\,, f \,, g}
P_{\ \ \ c\ a}^{+\ b\ g}\
\barR{f}{g}{c}(i\pi-\theta \,,  \xi_{+})\
\bar S_{f\ a}^{g\ d}(-2\theta+i\pi)\
\barR{k}{d}{f}(-\theta\,,  \xi_{+})\
P_{\ \ \ k\ a}^{+\ d\ j} \,. \non \\
\label{fusedRplus}
\ee
For the $NS$ case of the TIM, the nonzero matrix elements of $\tilde
{\bar R}^{+}$ are given by
\be
\fusedRplus{\sigma\ 0\ \ }{\quad \sigma}{\sigma\ 0\ \ } 
(\theta- i \pi\,, \xi_{+})
&=& \sqrt{2} \cosh{\theta\over 2} \,, \non \\
\fusedRplus{0\ \sigma\ \ }{\quad 0}{0\ \sigma'\ \ } (\theta- i \pi\,, \xi_{+})
&=&  {1\over 2\sqrt{2}}(\cosh \theta + \cos \xi_{+}) \cosh{\theta\over 2} \,.
\ee 

To summarize, the fused transfer matrix $\tilde{\bar t}$ is given by
(\ref{fusedtransfer})-(\ref{fusedmondoromy}), where the fused matrices
$\tilde{\bar S}$, $\tilde{\bar S}'$, $\tilde {\bar R}$ and 
$\tilde {\bar R}^{+}$ are given by (\ref{bulkSfused}), 
(\ref{bulkSfusedprime}), (\ref{fusedR}) and (\ref{fusedRplus}), 
respectively.
For the $NS$ case of the TIM, the nonzero matrix elements of the fused 
transfer matrix are as follows: for $N=$ even,
\be
\tilde{\bar t}^{\sigma_{1}\ 0\ \sigma_{2}\ \ldots \ 0\ 
\sigma_{{N\over 2}+1}}_{\sigma_{1}\ 0\ \sigma_{2}\ \ldots \ 0\ 
\sigma_{{N\over 2}+1}}(\theta | \theta_{1} \,, \ldots \,, \theta_{N})  
&=& 2 \cosh^{2}{\theta\over 2}\ (\prod \cosh) \,, \label{evenfusedresult} \\
\tilde{\bar t}^{0\ \sigma_{1}\ 0\ \sigma_{2} \ldots \sigma_{N\over 
2}\ 0}_{0\ \sigma_{1}\ 0\ \sigma_{2} \ldots \sigma_{N\over 
2}\ 0}(\theta | \theta_{1} \,, \ldots \,, \theta_{N}) 
&=& {1\over 2}(\cosh \theta + \cos \xi_{+})
(\cosh \theta + \cos \xi_{-})\cosh^{2}{\theta\over 2}\ (\prod \cosh) \,;
\non
\ee
and for $N=$ odd,
\be
\tilde{\bar t}^{\sigma_{1}\ 0\ \sigma_{2} \ldots \sigma_{N+1\over 2}
\ 0}_{\sigma_{1}\ 0\ \sigma_{2} \ldots \sigma_{N+1\over 2}\ 0}
(\theta | \theta_{1} \,, \ldots \,, \theta_{N}) 
&=& (\cosh \theta + \cos \xi_{-})\cosh^{2}{\theta\over 2}\
(\prod \cosh) \,, \non \\
\tilde{\bar t}^{0\ \sigma_{1}\ 0 \ldots 0\ \sigma_{N+1\over 2}}_{0\ 
\sigma_{1}\ 0 \ldots 0\ \sigma_{N+1\over 2}}
(\theta | \theta_{1} \,, \ldots \,, \theta_{N}) 
&=& (\cosh \theta + \cos \xi_{+})\cosh^{2}{\theta\over 2}\
(\prod \cosh) \,, \label{oddfusedresult}
\ee
where $(\prod \cosh)$ denotes
\be
(\prod \cosh) = \prod_{j=1}^{N}\cosh({1\over 2}(\theta-\theta_{j}))
\cosh({1\over 2}(\theta+\theta_{j})) \,.
\ee

As also discussed in Sec. \ref{sec:inversion}, for a given a transfer matrix 
(either fundamental 
$\bar t_{a'_{1} \ldots a'_{N+1}}^{a_{1} \ldots a_{N+1}}$
or fused 
$\tilde{\bar t}_{a'_{1} \ldots a'_{N+1}}^{a_{1} \ldots a_{N+1}}$),
it is convenient to define the following four ``sectors'':
\be
N=\mbox{ even } - \mbox{ Sector } I &:& 
a_{1}\,, a'_{1}\,, a_{N+1}\,, a'_{N+1}
\in \{-1 \,, +1\} \,, \non  \\
N=\mbox{ even } - \mbox{ Sector } II &:& 
a_{1} = a'_{1} = a_{N+1} = a'_{N+1} = 0 
\,, \non  \\
N=\mbox{ odd } - \mbox{ Sector } I &:& 
a_{1}\,, a'_{1} \in \{-1 \,, +1\}\,, 
\qquad a_{N+1} = a'_{N+1} = 0 \,, \non  \\
N=\mbox{ odd } - \mbox{ Sector } II &:& a_{1} = a'_{1} = 0 \,, 
\qquad a_{N+1} \,, a'_{N+1} \in \{-1 \,, +1\}  \,.
\label{sectorsapp}
\ee 
The results (\ref{evenfusedresult}), (\ref{oddfusedresult}) show that,
within each sector, the fused transfer matrix is proportional to the
adjacency-inclusive identity matrix,
\be
\tilde{\bar t}^{(\alpha)}(\theta| \theta_{1} \,, \ldots \,, \theta_{N}) 
= g^{(\alpha)}(\theta)\ \id^{(\alpha)} \,,
\label{fusionresult1}
\ee
where $\alpha$ runs over the four sectors (\ref{sectorsapp}), and 
$g^{(\alpha)}(\theta)$ is given by
\be
g^{(\alpha)}(\theta) =\cosh^{2}{\theta\over 2} 
\prod_{j=1}^{N}\cosh({1\over 2}(\theta-\theta_{j}))
\cosh({1\over 2}(\theta+\theta_{j})) \times
\left\{ \begin{array}{c}
          2 \\
	  {1\over 2}(\cosh \theta + \cos \xi_{+})(\cosh \theta + \cos 
	  \xi_{-})\\
	  (\cosh \theta + \cos \xi_{-})\\
	   (\cosh \theta + \cos \xi_{+})
	   \end{array}\right. \,, \label{fusionresult2}
\ee
respectively. This is a nontrivial property of the TIM. The supersymmetric
sinh-Gordon model enjoys \cite{AN} a similar property.

\subsection{Fusion formula and quantum determinants}

We now derive the important ``fusion formula,'' from which the
TIM inversion identity is obtained.  To this end, we first note that
$\tilde{\bar{\cal T}}$ (\ref{fusedcalT}) can be expressed as the
fusion of the corresponding fundamental quantities $\bar{\cal T}$
(\ref{calT}),
\be
\scriptTfused{a_{1}\ldots a_{N+1}}
{a''_{1}\,, b''_{1}\,, c''_{1}}{a'_{1}\ldots a'_{N+1}}
(\theta | \theta_{1} \,, \ldots \,, \theta_{N}) &=&
\sum_{b_{1}\,, \ldots \,, b_{N+1} \,, \atop f_{1} \,, g_{1}} \Big\{ 
P_{\ \ \ a''_{1}\ a'_{1}}^{+\ b''_{1}\ f_{1}}\
\scriptT{b_{1}\ldots b_{N+1}}{f_{1}}{a'_{1}\ldots a'_{N+1}}
(\theta | \theta_{1} \,, \ldots \,, \theta_{N})\
\bar S^{f_{1}\ g_{1}}_{a''_{1}\ b_{1}}(2\theta+ i\pi) \non \\
&\times& \scriptT{a_{1}\ldots a_{N+1}}{g_{1}}{b_{1}\ldots b_{N+1}}
(\theta+ i\pi | \theta_{1} \,, \ldots \,, \theta_{N})\
P_{\ \ \ a_{1}\ a''_{1}}^{+\ g_{1}\ c''_{1}} \Big\} \,.
\ee
We next observe that the reduced bulk $S$ matrix obeys
\be
\sum_{c}
\bar S^{a\ b}_{c\ d}(i\pi-2\theta)\ 
\bar S^{a\ b}_{c\ d'}(i\pi+2\theta) = \delta_{d\,, d'} 
\zeta_{d}(\theta) A_{a\ d} A_{b\ d} \,,
\ee
where, for the TIM, the scalar factor $\zeta_{d}(\theta)$ is given by
\be
\zeta_{0}(\theta)={1\over 2}\cosh \theta \,, \qquad
\zeta_{\pm 1}(\theta)=2\cosh \theta \,.
\ee
Using also Eqs. (\ref{fusedRplus}) and (\ref{calTYBE}),
we obtain the desired fusion formula \footnote{We save 
writing by suppressing the dependence of the transfer matrix, etc. 
on the inhomogeneity parameters $\theta_{1} \,, \ldots \,, \theta_{N}$.}
\be
\tilde{\bar t}^{a_{1} \ldots a_{N+1}}_{a'_{1} \ldots a'_{N+1}}(\theta) 
= \zeta_{a'_{1}}(\theta) 
\sum_{a''_{1}\,,  \ldots \,, a''_{N}} 
{\bar t}_{a'_{1} \ldots a'_{N}}^{a''_{1} \ldots a''_{N}}(\theta)\  
{\bar t}_{a''_{1} \ldots a''_{N}}^{a_{1} \ldots a_{N}}(\theta + i\pi) 
- \bar \Delta^{a_{1} \ldots a_{N+1}}_{a'_{1} \ldots a'_{N+1}}(\theta) \,,
\label{fusionformula}
\ee
where the quantum determinant \cite{IK, KS}
$\bar \Delta(\theta)$ of the transfer matrix is given by
\be
\bar \Delta^{a_{1} \ldots a_{N+1}}_{a'_{1} \ldots a'_{N+1}}(\theta)
&=&\sum_{f_{1}\,, \ldots \,, f_{N+1}\,, \atop
b\,, c\,, d\,, j\,, k} \Big\{
\qdetRplus{a_{1}\  d\ \ }{\quad f_{1}}{a'_{1}\ c\ \ }\ 
\qdetR{\ \ j\ a_{N+1}}{f_{N+1}\ \ \ }{\ \ b\ a'_{N+1}}\ 
P_{\ \ \ f_{1}\ a_{1}}^{-\ k\ \ d} \non \\
&\times&
\delta(\bar T(\theta))^{c\ f_{1}\ldots f_{N+1}}_{b\ a'_{1} \ldots 
a'_{N+1}}\
\delta(\hat {\bar T}(\theta))^{k\ a_{1}\ldots a_{N+1}}_{j\ f_{1} \ldots 
f_{N+1}} \Big\} \,,
\ee
where the quantum determinants of the monodromy matrices are defined by
\be
\delta(\bar T(\theta))^{c\ a_{1}\ldots a_{N+1}}_{b\ a'_{1} \ldots 
a'_{N+1}} &=&
\sum_{b_{1} \,, \ldots \,, b_{N+1}}
P_{\ \ \ a_{1}\ a'_{1}}^{-\ c\ \ b_{1}}\
\bar T^{b_{1}  \ldots b_{N+1}}_{a'_{1} \ldots a'_{N+1}}(\theta)\
\bar T^{a_{1}\ldots a_{N+1}}_{b_{1}  \ldots  b_{N+1}}
(\theta + i\pi)\
P_{\ \ \ a_{N+1}\ a'_{N+1}}^{-\ b_{N+1}\ b} \,, \non \\
\delta(\hat {\bar T}(\theta))^{c\ a_{1}\ldots a_{N+1}}_{b\ a'_{1} \ldots 
a'_{N+1}} &=&
\sum_{b_{1} \,, \ldots \,, b_{N+1}}
P_{\ \ \ a'_{1}\ a_{1}}^{-\ c\ \ b_{1}}\
\hat {\bar T}^{b_{1} \ldots b_{N+1}}_{a'_{1} \ldots a'_{N+1}}(\theta)\
\hat {\bar T}^{a_{1}\ldots a_{N+1}}_{b_{1}  \ldots  b_{N+1}}
(\theta + i\pi)\
P_{\ \ \ a'_{N+1}\ a_{N+1}}^{-\ b\ \ \ \ b_{N+1}} \,,
\ee
and the quantum determinants of the boundary matrices are defined by
\be
\qdetR{\ \ j\ k}{a\ \ \ }{\ \ b\ c}
&=& \sum_{d\,, f \,, g}
P_{\ \ \ a\ c}^{-\ b\ g}\
\barR{f}{g}{c}(\theta)\
\bar S_{a\ f}^{g\ d}(2\theta+i\pi)\
\barR{k}{d}{f}(\theta+i\pi)\
P_{\ \ \ a\ k}^{-\ d\ j} \,, \\
\qdetRplus{k\ j\ \ }{\quad a}{c\ b\ \ } 
&=& \sum_{d\,, f \,, g}
P_{\ \ \ c\ a}^{-\ b\ g}\
\barR{f}{g}{c}(i\pi-\theta \,,  \xi_{+})\
\bar S_{f\ a}^{g\ d}(-2\theta+i\pi)\
\barR{k}{d}{f}(-\theta\,,  \xi_{+})\
P_{\ \ \ k\ a}^{-\ d\ j} \,. \non 
\ee 

We now proceed to evaluate the quantum determinants for the TIM.
With the help of the identity
\be
\sum_{f \,, g}
P_{\ \ \ a\ c}^{-\ b\ g}\
\bar S_{g\ d}^{c\ f}(\theta)\ 
\bar S_{a\ f}^{g\ j}(\theta+ i\pi)\ 
P_{\ \ \ j\ d}^{-\ f\ k} = i \sqrt{2} \sinh{\theta\over 2}\ 
P_{\ \ \ a\ c}^{-\ b\ j}\
P_{\ \ \ j\ d}^{-\ c\ k} \,,
\ee
we find that the quantum determinants of the monodromy matrices are
given by
\be
\delta(\bar T(\theta))^{c\ a_{1}\ldots a_{N+1}}_{b\ a'_{1} \ldots 
a'_{N+1}} &=& (i \sqrt{2} \sinh{\theta\over 2})^{N}\ 
P_{\ \ \ a_{1}\ a'_{1}}^{-\ c\ \ a_{2}}\
P_{\ \ \ a_{2}\ a'_{2}}^{-\ a'_{1}\ \ a_{3}} \ldots
P_{\ \ \ a_{N+1}\ a'_{N+1}}^{-\ a'_{N}\ \ \ \ b} \,, \non \\
\delta(\hat {\bar T}(\theta))^{c\ a_{1}\ldots a_{N+1}}_{b\ a'_{1} \ldots 
a'_{N+1}} &=& (i \sqrt{2} \sinh{\theta\over 2})^{N}\ 
P_{\ \ \ a_{1}\ a'_{1}}^{-\ c\ \ a'_{2}}\
P_{\ \ \ a_{2}\ a'_{2}}^{-\ a_{1}\ \ a'_{3}} \ldots
P_{\ \ \ a_{N+1}\ a'_{N+1}}^{-\ a_{N}\ \ \ \ b} \,.
\ee
Moreover, for the NS case, the quantum determinants 
of the boundary matrices have the following nonzero 
matrix elements
\be
\qdetR{\ \ \ \ 0\ \sigma}{-\sigma\ \ \ }{\ \ \ \ 0\ \sigma}
&=& i \sqrt{2} \sinh{\theta\over 2} \,, \non \\
\qdetR{\ \ \sigma\ 0}{0\ \ \ }{\ \ \sigma'\ 0}&=&
\pm {i\over 2\sqrt{2}}(\cos \xi_{-} - \cosh \theta) 
\sinh{\theta\over 2} \,, \qquad \sigma = \pm \sigma' \,;
\ee 
and
\be
\qdetRplus{\sigma\  0\ \ \ }{\quad \ -\sigma}{\sigma\ 0\ \ \ }
&=& -i \sqrt{2} \sinh{\theta\over 2} \,, \non \\
\qdetRplus{0\ \sigma\ \ \ }{\quad \ 0}{0\ \sigma'\ \ \ }&=&
\pm {i\over 2\sqrt{2}}(\cosh \theta - \cos \xi_{+}) 
\sinh{\theta\over 2} \,, \qquad \sigma = \pm \sigma' \,.
\ee 
We conclude that the quantum determinant 
$\bar \Delta(\theta)$ for the NS case of the TIM is given by
\be
\bar \Delta^{(\alpha)}(\theta) = 
h^{(\alpha)}(\theta)\ \id^{(\alpha)} \,,
\label{qdetresult1}
\ee
where $\alpha$ runs over the four sectors (\ref{sectorsapp}), and 
$h^{(\alpha)}(\theta)$ is given by
\be
h^{(\alpha)}(\theta) =\sinh^{2}{\theta\over 2} 
\prod_{j=1}^{N}\sinh({1\over 2}(\theta-\theta_{j}))
\sinh({1\over 2}(\theta+\theta_{j})) \times
\left\{ \begin{array}{c}
          2 \\
	  {1\over 2}(\cosh \theta - \cos \xi_{+})(\cosh \theta - \cos 
	  \xi_{-})\\
	  (\cosh \theta - \cos \xi_{-})\\
	   (\cosh \theta - \cos \xi_{+})
	   \end{array}\right. \,, \label{qdetresult2}
\ee
respectively.  Substituting this result, together with the result
(\ref{fusionresult1}), (\ref{fusionresult2}) into the fusion formula
(\ref{fusionformula}), we finally arrive at the inversion identity
(\ref{inversionid1})-(\ref{inversionid3}).

\end{document}